\documentclass[twocolumn]{IEEEtran}
\usepackage{cite}
\usepackage{color,xcolor}
\usepackage{amsmath,amssymb,amsfonts}
\usepackage{algorithmic}
\usepackage{graphicx}
\usepackage{float}
\usepackage{epstopdf}
\usepackage{textcomp}
\usepackage{multirow}
\usepackage{booktabs}
\usepackage{subfigure}
\usepackage{stfloats}
\usepackage{color}
\usepackage{bm}
\def\BibTeX{{\rm B\kern-.05em{\sc i\kern-.025em b}\kern-.08em
    T\kern-.1667em\lower.7ex\hbox{E}\kern-.125emX}}

\begin{document}
\title{{ Single-Source SIE for Two-Dimensional Arbitrarily Connected Penetrable and PEC Objects with Nonconformal Meshes}}
\author{\IEEEauthorblockN{Zekun Zhu, \IEEEmembership{Graduate Student Member, IEEE}, Aipeng Sun, Xiaochao Zhou, \\ Shunchuan Yang, \IEEEmembership{Member, IEEE}, and Zhizhang (David) Chen, \IEEEmembership{Fellow, IEEE} \\}

\thanks{Manuscript received xxx; revised xxx.}
\thanks{This work was supported in part by the National Natural Science Foundation of China through Grant 61801010, Grant 62071125, Grant 61631002, and Fundamental Research Funds for the Central Universities. {\it {(Corresponding author: Shunchuan Yang)}}
	
Z. Zhu, A. Sun, X. Zhou are with the School of Electronic and Information Engineering, Beihang University, Beijing, 100083, China. (e-mail: zekunzhu@buaa.edu.cn, sap1997@163.com, zhouxiaochao@buaa.edu.cn)
	
S. Yang is with the Research Institute for Frontier Science and the School of Electronic and Information Engineering, Beihang University, Beijing, 100083, China. (e-mail: scyang@buaa.edu.cn)

Z. Chen was with College of Physics and Information Engineering, Fuzhou University and on leave from the Department of Electrical and Computer Engineering, Dalhousie University, Halifax, NS, Canada B3J 2X4. (e-mail: z.chen@dal.ca)
}
}

\maketitle

\begin{abstract}
 We proposed a simple and efficient modular single-source surface integral equation (SS-SIE) formulation for electromagnetic analysis of arbitrarily connected penetrable and perfectly electrical conductor (PEC) objects { in two-dimensional space}. In this formulation, a modular equivalent model for each penetrable object consisting of the composite structure is first independently constructed through replacing it by the background medium, no matter whether it is surrounded by the background medium, other media, or partially connected objects, and enforcing an equivalent electric current density on the boundary to remain fields in the exterior region unchanged. Then, by combining all the modular models and any possible PEC objects together, an equivalent model for the composite structure can be derived. The troublesome junction handling techniques are not needed and nonconformal meshes are supported. The proposed SS-SIE formulation is simple to implement, efficient, and flexible, which shows significant performance improvement in terms of CPU time compared with the original SS-SIE formulation and the Poggio-Miller-Chang-Harrington-Wu-Tsai (PMCHWT) formulation. Several numerical examples including the coated dielectric cuboid, the large lossy objects, the planar layered dielectric structure, and the partially connected dielectric and PEC structure are carried out to validate its accuracy, efficiency and robustness. 
\end{abstract}


\section{Introduction}
The method of moment (MOM) is widely used to solve various electromagnetic problems, such as scattering [\citen{MLFMA1}], radiation \cite{Ali1997hybrid}, parameter extraction in integrated circuits \cite{Zhu2005FastImp}, due to its unknowns residing on the boundaries of different piecewise homogeneous objects, which can significantly reduce the overall count of unknowns compared with the finite-difference time-domain (FDTD) method \cite{Taflove2005FDTD} and the finite element method (FEM) \cite{Jin2015FEM}.

To solve the challenging electromagnetic problems induced by arbitrarily connected objects, many efforts have been made in the MOM, such as the Poggio-Miller-Chang-Harrington-Wu-Tsai (PMCHWT) formulation \cite{Poggio1973Integral}, the combined tangential formulation (CTF) \cite{yla2005iterative}, the Müller formulation \cite{muller2013foundations}, the equivalence principle algorithms (EPAs) [\citen{Li2008multiscale}]-[\citen{Hu2010multilayered}], and so on. In those formulations, boundary conditions of the electric and magnetic fields are required to be enforced to guarantee correct solutions. Especially, junctions of multiple media intersection are required to be paid special attention to. { Although the management of junctions is simplified in the two-dimensional implementations, it still imposes challenging in the practical implementation.} Many techniques, such as the modified rooftop basis function [\citen{putnam1991Combined}]-[\citen{Carr2004procedure}], the current continuous boundary conditions [\citen{yla2005junctions}] [\citen{yla2005application}], are proposed to handle the boundary conditions on junctions. For those EPAs, fictitious boundaries are assumed not to be touched to avoid troublesome boundary issues [\citen{Li2008multiscale}] [\citen{Li2007wavefield}]. A macromodeling approach based on the PMCHWT formulation is recently proposed in \cite{Patel2020macromodeling} to model large inhomogeneous antenna arrays, in which fictitious interfaces are touched with each other through carefully handling various boundary conditions. However, those are bookkeeping formulations of the boundary conditions in different scenarios. A general formulation, which can handle arbitrarily connected objects, can hardly be obtained. In addition, both the equivalent electric and magnetic current densities are required in those formulations.

Various single-source (SS) formulations, such as the surface integral equation (SIE) with the differential surface admittance operator (DSAO) [\citen{Huynen2019broadband}]-[\citen{Huynen2020entire}], the single-source surface-volume-surface (SS-SVS) formulation [\citen{Menshov2014SVS-EFIE}], with only electric current density are developed. 
In \cite{Patel2014cable}, a single-source surface integral equation (SS-SIE) formulation is proposed to extract the electrical parameters of solid and hollow conductors. The media surrounding and inside the hollow conductors are assumed to be the same, and fictitious interfaces are not allowed to be touched in this formulation.
{ In [\citen{Zutter2005DSAO}] [\citen{Patel2018Macromodeling}], the SS-SIE formulations based on DSAO are proposed to solve the two- and three-dimensional homogeneous problems, where the composite structures with shared interfaces are not considered. In [\citen{Zhou2021embedded}] [\citen{Zhu2020vector SS-SIE}], a DSAO-based approach is proposed to model the objects embedded in multilayers. However, it is not applicable for the partially connected objects. In \cite{Zhou2021SS-SIE}, through carefully enforcing boundary conditions and eliminating all the interior unknowns, a SS-SIE formulation is proposed to model partially connected penetrable objects. However, it suffers from troublesome mathematical manipulations.}
 { In \cite{GIBC}, a single source formulation based on generalized impedance boundary condition (GIBC) is proposed. By mathematically eliminating the magnetic current density, only the integral formulation with only the electric current density is obtained. However, the GIBC formulations are not applicable for nonconformal meshes and require complex procedures to handle the boundary conditions.}

{ There are two types of nonconformal meshes in the SIE formulations: (1) nonconformation on the interfaces of different regions, (2) mismatching edges at adjacent elements of the same region. The first type exists when different regions are coupled without uniting meshes, whether it is a two-dimensional or three-dimensional problem. The second type only exists in the three-dimensional problems. There are several formulations proposed to handle the nonconformal meshes in PMCHWT formulation. In [\citen{NF-PMCHWT1}] [\citen{NF-PMCHWT2}], the segment-based in the two-dimensional space and the facet-based in the three dimensional basis functions are used to handle the two types of nonconformal meshes. In [\citen{NF-PMCHWT3}], the half-RWG function is used, and the second type of nonconformation in homogeneous region is handled.}

Domain decomposition methods (DDMs) can also efficiently model electrically large and multiscale composite structures. In [\citen{Lee2005non-overlapping}] [\citen{Peng2011integral}], the SIE-DDMs are proposed to solve the electromagnetic problems induced by complex composite structures. In those formulations, an approximation boundary condition, e.g., the transmission conditions, between adjacent domains is required to ensure electric current density to be continuous. However, those boundaries impose challenges for practical implementations if nonconformal meshes are used. 
{ In [\citen{DDM-DSAO1}] [\citen{DDM-DSAO2}], the SIE based on DSAO is proposed to model conductors with the rectangle or triangle cross sections in TM mode. To model the complex objects, the cross section has been decomposed into several triangles or rectangles. In [\citen{DDM-DSAO3}], the internal impedance of inhomogeneous conductors is calculated through the DSAO.}

\begin{figure}
	\centering
	\subfigure[]{
		\label{general_model_o} 
		\includegraphics[width=0.232\textwidth]{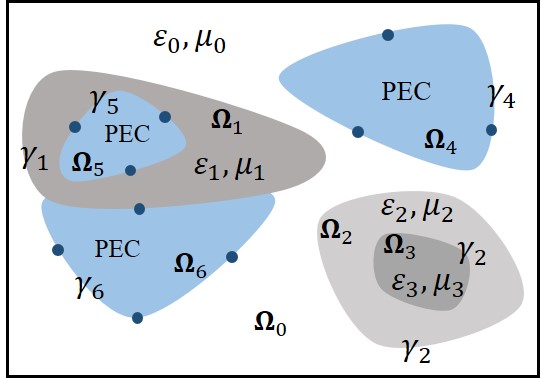}}
	\subfigure[]{
		\label{general_model_e} 
		\includegraphics[width=0.23\textwidth]{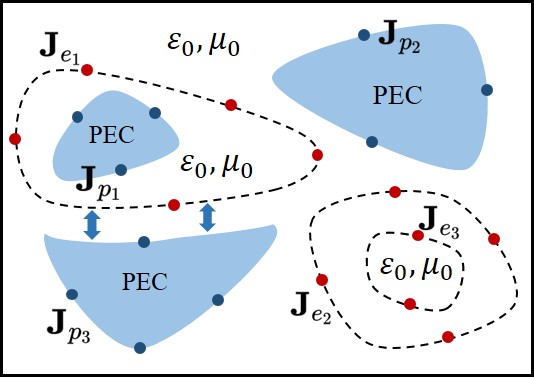}}
	\caption{(a) The general structure including arbitrarily connected penetrable and PEC objects, (b) the equivalent model with the physics and equivalent electric current densities on the boundaries of the composite structure.}
	\label{general_model} 
\end{figure}

In this paper, we proposed a generalized, simple and efficient modular SS-SIE formulation in { two-dimensional transverse magnetic (TM) mode} to model arbitrarily connected penetrable and perfectly electrical conductor (PEC) objects as shown in Fig. \ref{general_model}(a). In this formulation, an equivalent model with only the electric current density on the boundaries of the original structures is derived as shown in Fig. \ref{general_model}(b). First, through replacing each object by the background medium and enforcing an equivalent current density on the boundary, modular equivalent models for objects consisting of the composite structure are independently constructed no matter whether they are surrounded or partially connected with other objects. Only the surface equivalent electric current density is enforced on its corresponding boundary to keep fields unchanged in the exterior region by incorporating with the DSAO. Second, those modular equivalent models are combined into a single equivalent model for the composite structure, and the surface equivalent current density in the final equivalent model is equal to the summation of all the current densities on the interfaces of penetrable and PEC objects.

In \cite{Patel2017SS-SIE}, a modular approach to handle the composite structure is proposed. After each object is replaced by the equivalent current density, additional tangential boundary conditions of the magnetic fields are required. However, no such conditions are required and nonconformal meshes are supported in our proposed SS-SIE formulation. Compared with our previous work \cite{Zhou2021SS-SIE}, in which continuous boundary conditions are explicitly enforced on the shared interface before the surface equivalence theorem is applied, the proposed SS-SIE formulation in this paper can avoid the troublesome mathematical derivation and is suitable to model arbitrarily connected penetrable and PEC objects. Although slightly more unknowns compared with the SS-SIE formulation in \cite{Zhou2021SS-SIE} are required, significant performance improvement in terms of CPU time and high flexibility to model complex structures can be obtained as shown in the numerical examples. Therefore, it is much preferred in the practical simulations.

There are three obvious merits of the proposed SS-SIE formulation over other existing techniques.
\begin{enumerate}
	\item  Boundary conditions are implicitly enforced during the construction of the modular equivalent models for the penetrable objects, and no extra requirements are needed in the final equivalent model for the composite structures.  Therefore, the derivation of the SIE formulation for the general composite structures is significantly simplified and a general SS-SIE formulation for arbitrarily connected penetrable and PEC objects is obtained. 
	
	\item Nonconformal meshes are intrinsically supported since the equivalent model is modularly constructed and boundary conditions are automatically satisfied. { Especially for two-dimensional problems, only the first type non-conformation happens. Therefore, }each object can be discretized independently based on its own material parameters and geometric details. It is extremely flexible and useful to model multiscale and electrically large structures. Furthermore, much higher efficiency than that of the original SS-SIE formulation [\citen{Huynen2019broadband}]-[\citen{Huynen2020entire}] can be obtained through decomposing the large structures into small units.
	
	\item Only the single electric current density is required in our proposed SS-SIE formulation through incorporating with the DSAO. Shorter CPU time and less memory consumption are required compared with the dual source formulations, such as the PMCHWT formulation.
\end{enumerate}

This paper is organized as follows. In Section II, configurations and preliminary notations are defined. In addition, the proposed equivalent SS model for arbitrarily connected penetrable and PEC objects is presented. In Section III, detailed implementations for the proposed SS-SIE formulation are shown. In Section IV, we present how to solve the scattering problems with nonconformal meshes, near field calculation, and some discussion upon the proposed SS-SIE formulation are also presented. In Section V, several numerical examples are carried out to validate its accuracy, efficiency and robustness. At last, we draw some conclusions in Section VI.
 
\section{The Proposed Equivalent Model Based on The Surface Equivalence Theorem for Composite Objects}
\subsection{Configurations and Preliminary Notations}
\begin{figure}
	\centering
	\subfigure[]{
		\label{Die_Die} 
		\includegraphics[width=0.22\textwidth]{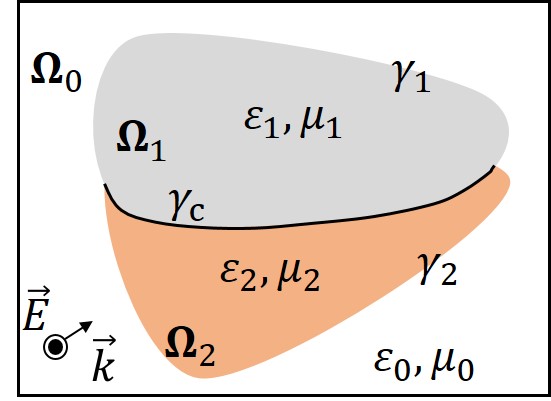}}
	\subfigure[]{
		\label{Die_PEC} 
		\includegraphics[width=0.216\textwidth]{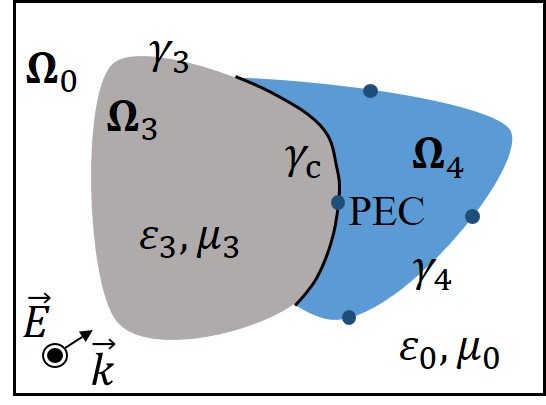}}
	\caption{(a) The partially connected structure for penetrable objects, (b) the penetrable and PEC partially connected objects.}
	\label{composite_model} 
\end{figure}
To make the derivation concise, two typical scenarios as shown in Fig. \ref{composite_model} are selected to illustrate the proposed SS-SIE formulation. Then, we demonstrate how the formulation can be extended to model arbitrarily connected penetrable and PEC objects. 

Fig. \ref{composite_model}(a) presents a composite structure including two partially connected penetrable objects denoted by $\Omega_{1}$ and $\Omega_{2}$. $\gamma_{1}$ and $\gamma_{2}$ are the boundaries of $\Omega_{1}$ and $\Omega_{2}$, respectively. $\gamma_{c}$ is the shared part of $\gamma_{1}$ and $\gamma_{2}$. The permittivity and permeability of $\Omega_{1}$ and $\Omega_{2}$ are $\varepsilon_1$, $\mu_1$ and $\varepsilon_2$, $\mu_2$, respectively. The permittivity $\varepsilon_{i}$ is $\varepsilon_{i}  = {\varepsilon _{0}}({\varepsilon _{r_i}} + j\sigma _i/(\varepsilon_0 \omega) )$ with $i=1,2$, where $\varepsilon_{r_i}$, $\sigma _i$ are the relative permittivity and the conductivity of the penetrable object, respectively, and $\omega$ is the angular frequency. The background medium is denoted by $\Omega_{0}$ with constant parameters $\varepsilon_0$, $\mu_0$. Similar notations are used for objects in Fig. \ref{composite_model}(b). $\Omega_{3}$ denotes the penetrable object with constant parameters of $\varepsilon_3, \mu_3$, and $\Omega_{4}$ for the PEC object. The hollow character with a subscript, e.g., $\mathbb{E}_i$, denotes a matrix associated to $\gamma_{i}$ and the bold character with a subscript, e.g., $\mathbf{E}_i$, denotes a column vector on $\gamma_{i}$. A quantity with $\,\widehat{}\,\,$, e.g., $\widehat{\mathbb{E}}$, is used for the equivalent model.

\subsection{The Equivalent Model with the Single Electric Current Density for A Penetrable Object}
\begin{figure}
	\centering
	\subfigure[]{
		\label{single_original} 
		\includegraphics[width=0.231\textwidth]{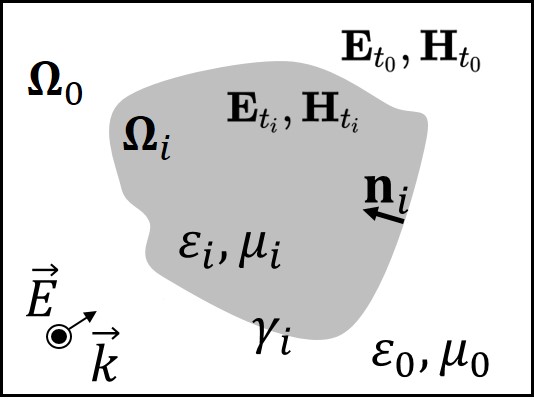}}
	\subfigure[]{
		\label{single_equivalent} 
		\includegraphics[width=0.23\textwidth]{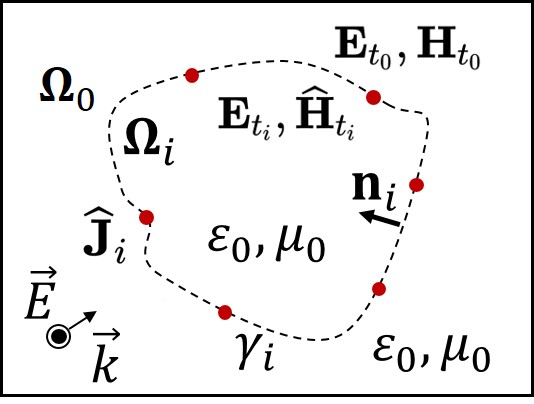}}
	\caption{(a) A penetrable object, (b) the equivalent model with the equivalent electric current densities for the penetrable object.}
	\label{single_model} 
\end{figure}
Let’s first consider a single penetrable object in Fig. \ref{single_model}(a). According to the surface equivalence theorem [\citen{LOVE}, Ch. 12], an appropriate equivalent electric and magnetic current density on an enclosed surface can reproduce exactly the same fields in the equivalent configuration as those in the original model. The surface equivalent electric and magnetic current densities can be expressed as
\begin{equation} \label{BDnH}
	\widehat{\mathbf{J}}_{i}(\mathbf{r})=\widehat{\mathbf{H}}_{t_{i}}(\mathbf{r})-\mathbf{H}_{t_{i}}(\mathbf{r}),
\end{equation}
\begin{equation}\label{BDnE}
	\widehat{\mathbf{M}}_{i}(\mathbf{r})=\widehat{\mathbf{E}}_{t_{i}}(\mathbf{r})-\mathbf{E}_{t_{i}}(\mathbf{r}),
\end{equation}
where $\mathbf{r} \in \gamma_{i}$, $\mathbf{E}_{t_{i}}(\mathbf{r}), \widehat{\mathbf{E}}_{t_{i}}(\mathbf{r}), \mathbf{H}_{t_{i}}(\mathbf{r}), \widehat{\mathbf{H}}_{t_{i}}(\mathbf{r})$ are the surface tangential electric and magnetic fields in the original and equivalent model, respectively. { The subscript $t$ denotes $\mathbf{n}\times$, and $\mathbf{E}_{t_{i}}(\mathbf{r})$ represents the $\mathbf{E}_z$ component in the z direction in TM mode.} The unit normal vector $\mathbf{n}_{i}$ pointing into the interior region of $\Omega_{i}$ is selected in this paper, and all the tangential fields are obtained through applying $\mathbf{n}_{i}\times$ operator to the corresponding fields.

When $\widehat{\mathbf{E}}_{t_{i}}(\mathbf{r}) = \widehat{\mathbf{H}}_{t_{i}}(\mathbf{r}) = \mathbf{0}$, it becomes the Love's equivalence theorem [\citen{LOVE}, Ch. 12]. Based on them, many SIE formulations [\citen{Poggio1973Integral}], [\citen{yla2005iterative}], [\citen{Patel2020macromodeling}] are derived to solve various electromagnetic problems. In this paper, we use them to derive the SS-SIE formulation.

Since fields in the equivalent model can be arbitrary, we enforce $\widehat{\mathbf{E}}_{t_{i}}(\mathbf{r}) = \mathbf{E}_{t_{i}}(\mathbf{r})$\cite{Zutter2005DSAO}, and (\ref{BDnE}) becomes
\begin{equation} \label{BDnE2}
	\widehat{\mathbf{M}}_{i}(\mathbf{r}) = {\mathbf{0}}.
\end{equation}
Then, the magnetic current density vanishes and only the electric current density exists in the equivalent model for a penetrable object as shown in Fig. \ref{single_model}(b). When $\mathbf{E}_{t_{i}}(\mathbf{r}) = {\mathbf{0}}$, (\ref{BDnH}) and (\ref{BDnE2}) correspond to the PEC objects, and $\mathbf{E}_{t_{i}}(\mathbf{r}) \neq {\mathbf{0}}$ is for the penetrable objects. Therefore, they are applicable for both the PEC and penetrable objects. In the next subsections, we will derive an equivalent model with only the electric current density for partially connected penetrable and PEC objects using (\ref{BDnH}) and (\ref{BDnE2}).

\subsection{The Proposed Equivalent Model with the Single Electric Current Density for Penetrable Composite Objects}
\begin{figure}
	\centering
	\subfigure[]{
		\label{DD_o} 
		\includegraphics[width=0.197\textwidth]{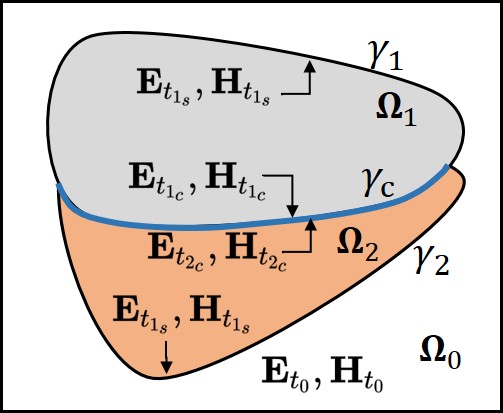}}
	\subfigure[]{
		\label{DD_e} 
		\includegraphics[width=0.2\textwidth]{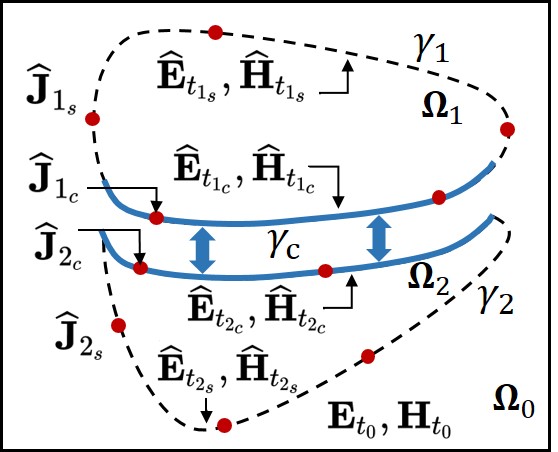}}
	\caption{(a) A composite structure with two penetrable objects, (b) the equivalent model for the composite structure and a small gap is added for better visualization.}
	\label{DD} 
\end{figure}
Let’s consider the scenario in Fig. \ref{DD}(a). According to the surface equivalence theorem [\citen{LOVE}, Ch. 12] and the SS formulation in Section II-B, to support exactly the same fields in the exterior region as those in the original configuration, the surface equivalent electric current densities are required to be enforced on $\gamma_{1}$ and $\gamma_{2}$, which can be expressed as
\begin{align}
\label{BD1i}	\widehat{\mathbf{J}}_{1_s}(\mathbf{r})&=\widehat{\mathbf{H}}_{t_{1_s}}(\mathbf{r}) - \mathbf{H}_{t_{1_s}}(\mathbf{r}), \mathbf{r} \in \gamma_{1},\mathbf{r} \notin \gamma_{c}, \\ 
 \label{BD2i}	\widehat{\mathbf{J}}_{2_s}(\mathbf{r})&=\widehat{\mathbf{H}}_{t_{2_s}}(\mathbf{r}) - \mathbf{H}_{t_{2_s}}(\mathbf{r}), \mathbf{r} \in \gamma_{2},\mathbf{r} \notin \gamma_{c},\\
\label{BDc}	\widehat{\mathbf{J}}_{c}(\mathbf{r})&=\widehat{\mathbf{H}}_{t_{2_c}}(\mathbf{r}) + \widehat{\mathbf{H}}_{t_{1_c}}(\mathbf{r}), \mathbf{r} \in \gamma_{c}, 
\end{align}
subject to the boundary conditions
\begin{align}
\label{BD1i_E}	{\widehat{\mathbf{E}}_{t_{1_s}}}(\mathbf{r}) &= {\mathbf{E}_{t_{1_s}}}(\mathbf{r}), \mathbf{r} \in \gamma_{1},\mathbf{r} \notin \gamma_{c},\\
\label{BD2i_E}	{\widehat{\mathbf{E}}}_{t_{2_s}}(\mathbf{r}) &= {\mathbf{E}_{t_{2_s}}}(\mathbf{r}), \mathbf{r} \in \gamma_{2},\mathbf{r} \notin \gamma_{c},\\
\label{BDc_E}	{\widehat{\mathbf{E}}_{t_{1_c}}}(\mathbf{r}) &= -{\widehat{\mathbf{E}}_{t_{2_c}}}(\mathbf{r}) = {{\mathbf{E}}_{t_{1_c}}}(\mathbf{r}) = -{{\mathbf{E}}_{t_{2_c}}}(\mathbf{r}), \mathbf{r} \in \gamma_{c},
\end{align}
where the subscript $1_s$ and $2_s$ denote that the quantities are defined on $\gamma_{1}$ and $\gamma_{2}$ except for $\gamma_{c}$, and the subscript $1_c$ and $2_c$ denote quantities defined on $\gamma_{c}$. { (\ref{BDc_E}) denotes that the tangential electric fields are continuous on the shared boundary, and they are unchanged in the original and equivalent configurations.} Since the unit normal vectors point into the interior region of each object, plus sign in (\ref{BDc}) and minus sign in (\ref{BDc_E}) should be used.

To obtain a general SIE formulation for arbitrary composite structures, fields on $\gamma_{c}$ need special treatments. By taking ${\mathbf{H}}_{t_{1_c}}(\mathbf{r}) = -{\mathbf{H}}_{t_{2_c}}(\mathbf{r})$ into consideration in the original model, (\ref{BDc}) can be modified as
\begin{equation} \label{J_seperated_PP}
	\begin{aligned}
		\widehat{\mathbf{J}}_{c}(\mathbf{r}) &=\widehat{\mathbf{H}}_{t_{2_c}}(\mathbf{r})+\widehat{\mathbf{H}}_{t_{1_c}}(\mathbf{r}) \\
		&=\underbrace{\left(\widehat{\mathbf{H}}_{t_{1_c}}(\mathbf{r})-\mathbf{H}_{t_{1_c}}(\mathbf{r})\right)}_{\widehat{\mathbf{J}}_{1_c}(\mathbf{r})}+\underbrace{\left(\widehat{\mathbf{H}}_{t_{2_c}}(\mathbf{r})-\mathbf{H}_{t_{2_c}}(\mathbf{r})\right)}_{\widehat{\mathbf{J}}_{2_c}(\mathbf{r})}.
	\end{aligned}
\end{equation} 
As shown in (\ref{J_seperated_PP}), $\widehat{\mathbf{J}}_{c}$ is split into two surface electric current densities, namely $\widehat{\mathbf{J}}_{1_c}$, $\widehat{\mathbf{J}}_{2_c}$ on $\gamma_{c}$. Therefore, we can assume two $\gamma_{c}$ exist. One $\gamma_{c}$ is on $\gamma_{1}$ and the other is on $\gamma_{2}$. $\widehat{\mathbf{J}}_{c}$ is the summation of two surface electric current densities, $\widehat{\mathbf{J}}_{1_c}$ on $\gamma_{c}$ of $\Omega_{1}$ and $\widehat{\mathbf{J}}_{2_c}$ on $\gamma_{c}$ of $\Omega_{2}$, as shown in Fig. \ref{DD}(b). Then, after combining ${\widehat{\mathbf{J}}}_{1_c}$ and ${\widehat{\mathbf{J}}}_{1_s}$, ${\widehat{\mathbf{J}}}_{2_c}$ and $\widehat{\mathbf{J}}_{2_s}$, (\ref{BD1i})-(\ref{BDc}) are rewritten as 
\begin{equation} \label{J1_Defined}
	\widehat{\mathbf{J}}_{1}(\mathbf{r})=\widehat{\mathbf{H}}_{t_{1}}(\mathbf{r})-\mathbf{H}_{t_{1}}(\mathbf{r}), \mathbf{r} \in \gamma_{1},
\end{equation} 
\begin{equation} \label{J2_Defined}
	\widehat{\mathbf{J}}_{2}(\mathbf{r})=\widehat{\mathbf{H}}_{t_{2}}(\mathbf{r})-\mathbf{H}_{t_{2}}(\mathbf{r}), \mathbf{r} \in \gamma_{2},
\end{equation} 
subject to the boundary conditions in (\ref{BD1i_E})-(\ref{BDc_E}). Up to this point, $\widehat{\mathbf{J}}_{1}$ and $\widehat{\mathbf{J}}_{2}$ are defined on the whole boundaries of $\Omega_{1}$ and $\Omega_{2}$, respectively. Therefore, the original partially connected objects are separated and the equivalent model for the composite structure can be constructed through combining the modular models for each penetrable object.

As shown in the previous subsection, (\ref{J1_Defined}) and (\ref{J2_Defined}) are the equivalent electric current densities enforced on the boundaries of $\Omega_{1}$ and $\Omega_{2}$, respectively, when they are replaced by the background medium. It is also true for arbitrarily connected penetrable objects. Therefore, we can obtain the following procedure to derive the equivalent model for any penetrable composite objects.
\begin{enumerate}
	\item A modular equivalent model for each penetrable object of the composite structure in Fig. \ref{DD}(a) is first derived by using (\ref{BDnH}) and (\ref{BDnE2}), no matter whether the object is surrounded by the background medium, other media, or connected with other objects as shown in Fig. \ref{general_model}(a).
	\item All the modular equivalent models are combined together to construct the equivalent model for the composite structure. The equivalent current density in the final equivalent model is equal to the summation of all the modular equivalent current densities.
\end{enumerate}

{ In traditional PMCHWT formulation, the current densities are treated as one set of variables on the shared interfaces through the boundary conditions. In the proposed SS-SIE formulation, the modular electric current densities are used as independent variables, which is similar to the method in [\citen{ZYZ-PMCHWT}]. Therefore, nonconformal meshes are supported. These schemes provide flexibility in the modular design, but the additional unknowns are required at the interfaces.}

\subsection{The Proposed Equivalent Model with the Single Electric Current Density for Penetrable and PEC Connected Objects}
\begin{figure}
	\centering
	\subfigure[]{
		\label{DP_o} 
		\includegraphics[width=0.232\textwidth]{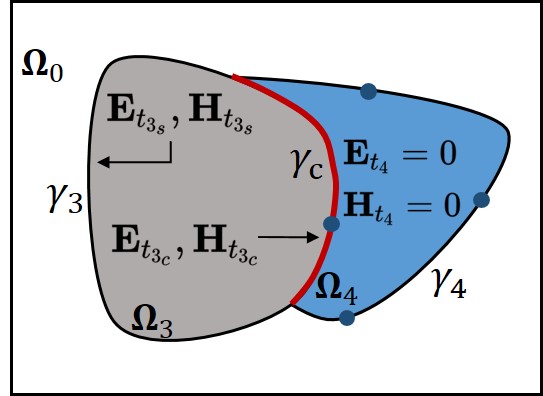}}
	\subfigure[]{
		\label{DP_e} 
		\includegraphics[width=0.228\textwidth]{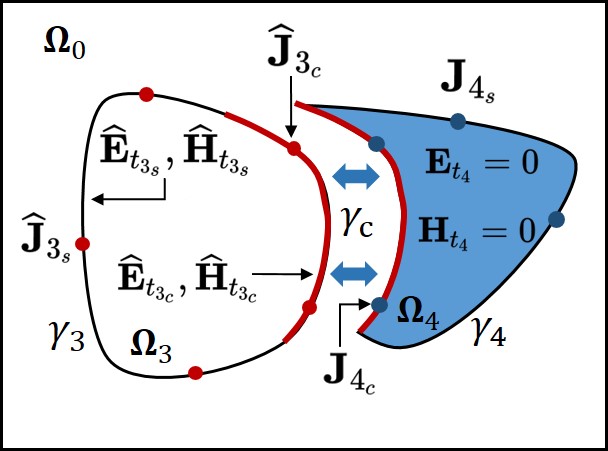}}
	\caption{(a) A composite structure consisting of a penetrable and PEC partially connected object, (b) the equivalent model for the composite structures and a small gap is added for better visualization.}
	\label{DP} 
\end{figure}
Let’s consider the partially connected PEC and penetrable objects in Fig. \ref{composite_model}(b). Compared with the previous scenario in Fig. \ref{composite_model}(a), one physics rather than the equivalent electric current density ${\mathbf{J}}_{4}$ exists on the surface of the PEC object. Therefore, the surface equivalent theorem is not needed to be applied to those PEC objects. The electric current density on the boundary can be expressed as 
\begin{align}
\label{BDPECi}	\mathbf{J}_{{4_s}}(\mathbf{r})&=\mathbf{H}_{t_{0}}(\mathbf{r}), \mathbf{r} \in \gamma_{4},\mathbf{r} \notin \gamma_{c},\\
\label{BDPECc}	\mathbf{J}_{{4_c}}(\mathbf{r})&=\mathbf{H}_{t_{3_c}}(\mathbf{r}), \mathbf{r} \in \gamma_{c},
\end{align}
where $\mathbf{H}_{t_{0}}(\mathbf{r})$ is the tangential magnetic fields on $\gamma_{4}$ except for $\gamma_{c}$ and $\mathbf{H}_{t_{3_c}}(\mathbf{r})$ denotes the tangential magnetic fields in $\Omega_{3}$ on $\gamma_{c}$.

For the penetrable object $\Omega_{3}$, the modular equivalent model can be derived through the procedure in the previous subsection. Similarly, the penetrable object is replaced by the background medium, and the electric current density in the final equivalent configuration in Fig. \ref{DP}(b) can be expressed as
\begin{align}
\label{BD3i}  \widehat{\mathbf{J}}_{{3_s}}(\mathbf{r})&=\widehat{\mathbf{H}}_{t_{3_s}}(\mathbf{r})-\mathbf{H}_{t_{3_s}}(\mathbf{r}), \mathbf{r} \in \gamma_{3},\mathbf{r} \notin \gamma_{c},\\
\label{BDc_PD}	\widehat{\mathbf{J}}_{{c}}(\mathbf{r})\  &=\widehat{\mathbf{H}}_{t_{3_c}}(\mathbf{r}), \mathbf{r} \in \gamma_{c},
\end{align} 
where the subscript $3_s$ denotes the quantities are defined on $\gamma_{3}$ except for $\gamma_{c}$. 
Similarly, (\ref{BDc_PD}) can be modified as
\begin{equation}\label{J_seperated_PD}
	\begin{aligned}
		\widehat{\mathbf{J}}_{c}(\mathbf{r}) &=\widehat{\mathbf{H}}_{t_{3_c}}(\mathbf{r}) \\
		&=\widehat{\mathbf{H}}_{t_{3_c}}(\mathbf{r})-\mathbf{H}_{t_{3_c}}(\mathbf{r})+\mathbf{H}_{t_{3_c}}(\mathbf{r}) \\
		&=\underbrace{\left(\widehat{\mathbf{H}}_{t_{3_c}}(\mathbf{r})-\mathbf{H}_{t_{3_c}}(\mathbf{r})\right)}_{\widehat{\mathbf{J}}_{3_c}(\mathbf{r})}+\underbrace{\mathbf{H}_{t_{3_c}}(\mathbf{r})}_{{\mathbf{J}}_{4_c}(\mathbf{r})}.
	\end{aligned}
\end{equation}
Therefore, the summation of $\widehat{\mathbf{J}}_{{3_s}}$ in (\ref{BD3i}) and $\widehat{\mathbf{J}}_{{3_c}}$ in (\ref{J_seperated_PD}) is the modular equivalent electric current density on $\gamma_{3}$. The summation of ${\mathbf{J}}_{{4_s}}$ in (\ref{BDPECi}) and ${\mathbf{J}}_{{4_c}}$ in (\ref{J_seperated_PD}) is equal to the physics electric current density on $\gamma_{4}$. It is also true for the arbitrarily connected penetrable and PEC objects. The equivalent model for the composite structure can be also obtained through the similar procedure in previous subsection.

In the following section, we present detailed implementations of the proposed equivalent model in the MOM and how it can support nonconformal meshes.

\section{Detailed Implementations of the Proposed SS-SIE Formulation}
\subsection{The Modular Equivalent Model for Each Penetrable Object}
In this paper, we consider the TM mode to demonstrate the implementations of the proposed SS-SIE formulation. However, it can also be used to solve more general vector EM problems. In the appendix, a numerical example is provided to demonstrate the applicability of the proposed SS-SIE formulation to solve the vector TE electromagnetic problems. { We then present the implementations of the proposed formulation, and the readers can refer to [\citen{Zhou2021embedded}] [\citen{Zhu2020vector SS-SIE}] for more implementation details.}

Without loss of generality, there are no sources inside the penetrable region $\Omega_i$. The electric fields on $\gamma_{i}$ can be expressed as
\begin{align} 
	T {E}_{i}(\mathbf{r})=\oint_{\gamma_{i}}&\bigg{[} G_{i}\left(\mathbf{r}, \mathbf{r}^{\prime}\right) {\frac{\partial {E}_{i}\left(\mathbf{r}^{\prime}\right)}{\partial {n}^{\prime}}}\notag\\
	\label{Contour_Integral}-&\left.\frac{\partial G_{i}\left(\mathbf{r}, \mathbf{r}^{\prime}\right)}{\partial {n}^{\prime}} {E}_{i}\left(\mathbf{r}^{\prime}\right)\right] d \mathbf{r}^{\prime},
\end{align}
where $T=1/2$ when the source point $\mathbf{r}^{\prime}$ and the observation point $\mathbf{r}$ are located on the same boundary, otherwise, $T=1$. $G_i\left(\mathbf{r}, \mathbf{r}^{\prime}\right)$ is the Green's function with $G_{i}\left(\mathbf{r}, \mathbf{r}^{\prime}\right)=-j H_{0}^{(2)}\left(k_{i}|\bm{\rho}|\right) / 4$, $j=\sqrt{-1}$, $\bm{\rho} = \mathbf{r}-\mathbf{r}^{\prime}$, where $k_i$ is the wavenumber inside the object, $H_{0}^{(2)}\left(\cdot\right)$ is the {\it{zero}}th-order of Hankel function of the second kind. According to the Poincare-Steklov operator [\citen{Zutter2005DSAO}], we have 
\begin{equation} \label{H=derivativesEn}
	{H}_{t_i}(\mathbf{r})=\left.\frac{1}{j \omega \mu_{i}} \frac{\partial {E}_{i}(\mathbf{r})}{\partial {n}}\right|_{\mathbf{r} \in \gamma_{i}},
\end{equation}
where $\mu_{i}$ is the permeability of the penetrable object. By substituting (\ref{H=derivativesEn}) into (\ref{Contour_Integral}), we have
\begin{align}
		T {E}_{i}(\mathbf{r})=\oint_{\gamma_{i}}&\bigg{[}j \omega \mu_{i} G_{i}\left(\mathbf{r}, \mathbf{r}^{\prime}\right) {H}_{t_i}(\mathbf{r'})\notag\\
		\label{Contour_Integral2} -&\left.\frac{\partial G_{i}\left(\mathbf{r}, \mathbf{r}^{\prime}\right)}{\partial {n}^{\prime}} {E}_{i}\left(\mathbf{r}^{\prime}\right)\right] d \mathbf{r}^{\prime},
\end{align}
After $\gamma_i$  is discretized into $m_i$ segments, the tangential electric and magnetic fields can be expanded as
\begin{equation}\label{E_expansion}
	{E}_{i}(\mathbf{r})=\sum_{n=1}^{m_{i}} e_{n} f_{n}(\mathbf{r}),\ \  {H}_{t_i}(\mathbf{r})=\sum_{n=1}^{m_{i}} h_{n} f_{n}(\mathbf{r}),
\end{equation}
where $f_{n}$ is the pulse basis function, $e_n$ and $h_n$ are the expansion coefficients, respectively. We collect all the expansion coefficients into column vectors as $\mathbf{E}_{i}$ and $\mathbf{H}_{i}$, which can be expressed as 
	\begin{align}
		\mathbf{E}_{i}&=\left[\begin{array}{llll}
			e_{1} & e_{2} & \cdots & e_{m_{i}}
		\end{array}\right]^{T}, \notag\\
		\label{co_EandH}\mathbf{H}_{i}&=\left[\begin{array}{llll}
			h_{1} & h_{2} & \cdots & h_{m_{i}}
		\end{array}\right]^{T}.
	\end{align}
By substituting (\ref{E_expansion}) into (\ref{Contour_Integral2}) and using the Galerkin scheme on (\ref{Contour_Integral2}), we get the following matrix equation
\begin{equation} \label{MatrixEqu_EH}
	T\, \mathbb{L}_{i} \mathbf{E}_{i}=\mathbb{P}_{i} \mathbf{H}_{i}+\mathbb{U}_{i} \mathbf{E}_{i},
\end{equation}
where $\mathbb{L}_{i}$ is a diagonal matrix, in which entities are equal to the length of segments. The entities of matrix $\mathbb{P}_{i}$ and $\mathbb{U}_{i}$ are given by
\begin{align}
\label{K_operator}	\left[\mathbb{U}_{i}\right]_{m, n}&=\int_{\gamma_{i_{m}}} \int_{\gamma_{i_{n}}} k_{i} \frac{\bm{\rho} \cdot\mathbf{ n}^{\prime}}{|\bm{\rho}|} G_{i}\left(\mathbf{r}, \mathbf{r}^{\prime}\right) d \mathbf{r}^{\prime} d \mathbf{r},\\
\label{L_operator}	\left[\mathbb{P}_{i}\right]_{m, n}&=\int_{\gamma_{i_{m}}} \int_{\gamma_{i_{n}}} j \omega \mu_{i} G_{i}\left(\mathbf{r}, \mathbf{r}^{\prime}\right) d \mathbf{r}^{\prime} d \mathbf{r}.
\end{align}

To handle the nearly singular and singular integrals  in (\ref{K_operator}) and (\ref{L_operator}), the approach in \cite{Zhu2020vector SS-SIE} is used. Interested readers are referred to it for more details.

Then, by moving the second terms on the right hand side (RHS) of (\ref{MatrixEqu_EH}) to its left hand side (LHS) and inverting the square matrix $\mathbb{P}_{i}$, we obtain 
\begin{equation}\label{Y_Defined}
	\mathbf{H}_{i}=\underbrace{\left[\mathbb{P}_{i}\right]^{-1}\left(T\, \mathbb{L}_{i}-\mathbb{U}_{i}\right)}_{\mathbb{Y}_{i}} \mathbf{E}_{i},
\end{equation}
where $\mathbb{Y}_{i}$ is the surface admittance operator (SAO) \cite{Zutter2005DSAO}.

By applying the surface equivalence theorem to the penetrable object, it is replaced by the background medium and an equivalent surface electric current density is enforced on $\gamma_{i}$ with $\widehat{\mathbf{E}}_{t_{i}}(\mathbf{r}) = \mathbf{E}_{t_{i}}(\mathbf{r})$. In this configuration, the tangential magnetic fields can be expanded as
\begin{equation} \label{He_expansion}
	\widehat{{H}}_{t_i}(\mathbf{r})=\sum_{n=1}^{m_{i}} \widehat{h}_{n} f_{n}(\mathbf{r}).
\end{equation}
We collect all the expansion coefficients into a column vector $\widehat{\mathbf{H}}_{i}$, which can be expressed as
\begin{equation} \label{co_Hhat}
	\begin{array}{l}
		\widehat {\mathbf{H}}_{i}=\left[\begin{array}{llll}
			\widehat{h}_{1} & \widehat{h}_{2} & \cdots & \widehat{h}_{m_{i}}
		\end{array}\right]^{T}. \\	
	\end{array}
\end{equation}

With the similar procedure in the original model, we obtain
\begin{equation}\label{Ye_Defined}
	\widehat{\mathbf{H}}_{i}=\underbrace{\left[\widehat{\mathbb{P}}_{i}\right]^{-1}\left(T\, \mathbb{L}_{i}-\widehat{\mathbb{U}}_{i}\right)}_{\widehat{\mathbb{Y}}_{i}} \mathbf{E}_{i}.
\end{equation}
The equivalent current density can be expanded by the pulse basis function, and the expansion coefficients are written into a column vector  $\mathbf{J}_{e_i}$ as
\begin{equation}\label{J_co}
	\mathbf{J}_{{e_i}}=\left[\begin{array}{llll}
		j_{{1}} & j_{{2}} & \cdots & j_{{m_i}}
	\end{array}\right]^{T}.
\end{equation}
By substituting (\ref{Y_Defined}) and (\ref{Ye_Defined}) into (\ref{BDnH}),  $\mathbf{J}_{e_i}$ can be expressed as
\begin{equation}\label{Ys_Defined}
	\mathbf{J}_{e_i}=\widehat{\mathbf{H}}_{i}-{\mathbf{H}_{i}}=\mathbb{Y}_{s_{i}} \mathbf{E}_{i},
\end{equation}
where $\mathbb{Y}_{s_{i}}$ is the DSAO \cite{Zutter2005DSAO} given by
\begin{align}
	\mathbb{Y}_{s_{i}}=\widehat{\mathbb{Y}}_{i}-\mathbb{Y}_{i}&=\left[\widehat{\mathbb{P}}_{i}\right]^{-1}\left(T \, \mathbb{L}_{i}-\widehat{\mathbb{U}}_{i}\right)\notag\\[0.25em]
\label{Ys_E_H_He}	&-\left[\mathbb{P}_{i}\right]^{-1}\left(T \, \mathbb{L}_{i}-\mathbb{U}_{i}\right).
\end{align}
\newcounter{TempEqCnt} 
\setcounter{TempEqCnt}{\value{equation}} 
\setcounter{equation}{40} 
\begin{figure*}[hb] 
	\hrulefill  
	\begin{equation}\label{EFIE_P}
		\mathbb{P}=\left[\begin{array}{l|l}
			{\overbrace{  \begin{array}{lll}
						\mathbb{P}_{(1,1)} & \ \ \ \;\cdots & \mathbb{P}_{(1,M)} \\
						\;\;\;\;\vdots & \ \ \ \;\ddots & \;\;\;\;\vdots \\
						\mathbb{P}_{(M,1)} &\ \ \ \;\cdots & \mathbb{P}_{(M,M)}
				\end{array} } ^{\mathbb{P} _{(e,e)}}} &  {\overbrace{ \begin{array}{lll}
						\mathbb{P}_{(1,M+1)} &  \ \ \ \;\cdots & \mathbb{P}_{(1,M+N)} \\
						\;\;\;\;\;\vdots &  \ \ \ \;\ddots & \;\;\;\;\;\vdots \\
						\mathbb{P}_{(M,M+1)} & \ \ \ \;\cdots & \mathbb{P}_{(M,M+N)} 
				\end{array}}  ^{\mathbb{P} _{(e,p)}}}  \\
			\hline {\underbrace{ \begin{array}{lll}
						\mathbb{P}_{(M+1,1)} & \cdots & \mathbb{P}_{(M+1,M)} \\
						\;\;\;\;\vdots & \ddots & \;\;\;\;\vdots \\
						\mathbb{P}_{(M+N,1)} &\cdots & \mathbb{P}_{(M+N,M)}
				\end{array}} _{\mathbb{P} _{(p,e)}}} &{\underbrace{ \begin{array}{lll}
						\mathbb{P}_{(M+1,M+1)} & \cdots & \mathbb{P}_{(M+1,M+N)} \\
						\;\;\;\;\vdots & \ddots & \;\;\;\;\vdots \\
						\mathbb{P}_{(M+N,M+1)} &\cdots & \mathbb{P}_{(M+N,M+N)}
				\end{array} } _{\mathbb{P} _{(p,p)}}}
		\end{array}\right]
	\end{equation}
\end{figure*}
\setcounter{equation}{\value{TempEqCnt}}

\subsection{Expansion of the Electric Current Density for PEC Objects}
If the object in $\Omega_{i}$ is PEC, the surface electric current density on the interface of PEC object can be expanded as
\begin{equation} \label{Jpec_expansion}
	{J}_{p_i}(\mathbf{r})=\sum_{n=1}^{m_{i}} j_{n} f_{n}(\mathbf{r}),
\end{equation}
where ${J}_{p_i}$ is the physics current densities on the boundary of $\Omega_{i}$, and $j_n$ is the corresponding expansion coefficient. Then, we collect all the expansion coefficients into a column vector $\mathbf{J}_{{p_i}}$ as
\begin{equation} \label{Jpec_co}
	\mathbf{J}_{{p_i}}=\left[\begin{array}{llll}
		j_{{1}} & j_{{2}} & \cdots & j_{{m_i}}
	\end{array}\right]^{T}.
\end{equation}

Once all the penetrable objects are replaced by the background medium, the equivalent current densities in (\ref{J_co}) and the physics current densities in (\ref{Jpec_co}) are obtained. Then, we can construct the equivalent model for the composite structures. 

\subsection{The Electric Current Density in the Equivalent Model for the Composite Structure}
As stated in the previous section, the electric current density in the final equivalent model is obtained through collecting all the modular equivalent electric current densities in (\ref{J_co}) and the physics current density in (\ref{Jpec_co}), which is expressed as
\begin{equation} \label{Jall}
	\mathbf{J}=\left[\begin{array}{l}
		\mathbf{J}_{e} \\
		\mathbf{J}_{p}
	\end{array}\right],
\end{equation}
Assume that there are $M$ penetrable objects and $N$ PEC objects, and the unknowns on the boundaries of penetrable and PEC objects are collected together for a more intuitive representation. $\mathbf{J}_e$ and $\mathbf{J}_p$ are expressed as 
\begin{align} 
\label{Je}	\mathbf{J}_{e} &=\left[\begin{array}{lll}
		\mathbf{J}_{e_{1}} & \cdots & \mathbf{J}_{e_{M}}
	\end{array}\right]^{T}, \\
\label{Jp}	\mathbf{J}_{p} &=\left[\begin{array}{lll}
	\mathbf{J}_{p_{1}} & \cdots & \mathbf{J}_{p_{N}}
	\end{array}\right]^{T}.
\end{align}

When the electric current density in the final equivalent model is defined in (\ref{Jall}), we can solve the electromagnetic problem through combining the electric field integral equation (EFIE) in the exterior region.

\section{Scattering Modeling}
\subsection{Scattering Modeling}
When a plane wave incidents from the exterior region, the total electric fields in TM mode can be expressed as
\begin{equation}\label{EFIE}
	{E}(\mathbf{r})=-j \omega \mu_{0} \int_{\gamma} G_{0}\left(\mathbf{r}, \mathbf{r}^{\prime}\right) {J}\left(\mathbf{r}^{\prime}\right) d \mathbf{r}^{\prime}+{E}^{i n c}(\mathbf{r}),
\end{equation}
where ${E}^{inc}$ denotes the plane wave, $\gamma$ is the union of all enclosed boundaries, and ${J}$ is the electric current densities consisting of the equivalent and physics current densities. When the observation points are fixed on the boundaries, by substituting (\ref{Jall}) into (\ref{EFIE}) and testing it on $\gamma$, we obtain the following matrix equation
\begin{equation}\label{EFIE_Mat}
	\left[\begin{array}{l}
		{ \mathbb{L}}\mathbf{E} \\
		\ \mathbf{0}
	\end{array}\right]=\mathbb{P}\left[\begin{array}{l}
		\mathbf{J}_{e} \\
		\mathbf{J}_{p}
	\end{array}\right]+\mathbf{E}^{inc},
\end{equation}
where the entities of $\mathbf{E}$ with dimension of $\sum\nolimits_{i = 1}^M {{m_i}}$ are the expansion coefficients of the tangential electric fields  and the entities of $\mathbf{0}$ with dimension of $\sum\nolimits_{i = M + 1}^{M + N} {{m_i}}$ are all zeros. $\mathbb{L}$ is a diagonal matrix, which is given by 
\begin{equation}\label{EFIE_L}
	{ \mathbb{L}}=\left[
	 \begin{array}{ccc}
				\mathbb{L}_{1} & & \\
			&\ddots & 	\\
				&  & \mathbb{L}_{M}
		\end{array} 
	\right].
\end{equation}
$\mathbb{P}$ is given by (\ref{EFIE_P}) at the bottom of this page. In (\ref{EFIE_P}), the subscript $(i,j)$ denotes the testing and source boundaries are $\gamma_{j}$, $\gamma_{i}$, respectively.

{ It should be noted that $\mathbb{L}$ is a diagonal matrix, because we treat $\mathbf{E}$ on the shared boundary as two independent variables. Although they are placed on the same location, only the main diagonal entity of $\mathbb{L}$ is nonzero. As shown in (\ref{EFIE}), one electric field is obtained when a certain point $\mathbf{r}$ is fixed, so the boundary condition in (\ref{BDc_E}) is satisfied. }

By moving the first term on the RHS of (\ref{EFIE_Mat}) to its LHS, and inverting the square coefficient matrix, we obtain 
\setcounter{equation}{41}
\begin{equation} \label{EFIE_general_Mat}
	\left[\begin{array}{c}
		\mathbf{E} \\
		\mathbf{J}_{p}
	\end{array}\right]=\left[\begin{array}{cc}
		{ \mathbb{L}}-\mathbb{P}_{(e,e)} \mathbb{Y}_{s} & -\mathbb{P}_{(e,p)} \\
		-\mathbb{P}_{(p,e)}{ \mathbb{Y}_{s}} & -\mathbb{P}_{(p,p)}
	\end{array}\right]^{-1} \mathbf{E}^{i n c},
\end{equation}
where $\mathbb{Y}_{s}$ is the DSAO of the composite structure assembling from the modular DSAO for all penetrable objects, which is expressed as 
\begin{equation} \label{EFIE_Ys}
	\mathbb{Y}_{s}=\left[\begin{array}{llll}
		\mathbb{Y}_{s_1} & & & \\
		& \mathbb{Y}_{s_2} & & \\
		& & \ddots & \\
		& & & \mathbb{Y}_{s_M}
	\end{array}\right].
\end{equation}

Through solving (\ref{EFIE_general_Mat}), the tangential electric fields and the physics electric current densities can be obtained. Then, other interested parameters, such as near fields, the radar cross section (RCS), can be calculated. In the following subsection, we introduce how near fields can be calculated through the tangential electric fields and the physics electric current density.

\subsection{Near Field Calculation}
There are three scenarios for near field calculation. First, the electric fields on the boundary of each dielectric objects have been computed through (\ref{EFIE_general_Mat}). Second, after all equivalent and physics current densities are calculated from (\ref{EFIE_general_Mat}), the electric fields in the exterior regions of the original structures can be calculated according to ($\ref{EFIE}$). Third, through substituting $\mathbf{E}_i$ into (\ref{Y_Defined}), $\mathbf{H}_i$ can be obtained, and then the electric fields in the interior regions of dielectric objects can be calculated through (\ref{Contour_Integral2}), where the constant parameters in the original structures should be used.

%

\subsection{Discussion}
There are several obvious merits compared with other existing techniques to handle complex composite structures in the proposed SS-SIE formulation.
First, each object is modularly modeled to derive the equivalent current density. It is easy to implement the proposed SS-SIE formulation.
Second, when the equivalent model for the composite structures is constructed, the equivalent current density is the summation of each modular current densities enforced on the penetrable object. Furthermore, they are treated as independent quantities. There are no requirements upon meshes on the connected interfaces, which implies that nonconformal meshes are intrinsically supported. As other nonconformal DDMs usually require to enforce the boundary conditions, it leads to challenges for implementation in the practical engineering problems. However, since the boundary conditions for the electric fields are automatically satisfied, no additional requirements in the proposed SS-SIE formulation are needed. It is extremely easy to implement the proposed SS-SIE formulation and useful to model multiscale and electrically large structures. Third, since the DSAO is incorporated into the proposed SS-SIE formulation, only the electric current density is required. 

One main bottleneck for the original SS-SIE formulations incorporated the DSAO to model multiscale and electrically large objects is that the construction of the DSAO is computationally intensive since matrix inversion is required as shown in (\ref{Ys_E_H_He}). Therefore, those formulations are mainly used to model small periodic and quasi-periodic scatters in [\citen{Patel2020macromodeling}] [\citen{Zhou2021embedded}]. However, the proposed SS-SIE formulation in this paper does not suffer from such issue as we can divide those structures into small units with the same constant parameters as those of the original object. We only needs to calculate the DSAO for small units. Therefore, its construction can be significantly accelerated. In addition, the large structures can be partitioned into many identical units. Since only one DSAO is required for each type of units, the efficiency can be further improved. Although there are slightly more unknowns required in the proposed SS-SIE formulation compared with those in [\citen{Zhou2021embedded}], the overhead can be ignored for challenging electromagnetic simulations compared with the overall computational cost in terms of CPU time and memory consumption. Significant less CPU time and high flexibility to model complex structures can be obtained as shown in the numerical examples in the later section.

\section{Numerical Results And Discussion}
Our in-house codes are developed in Matlab and all simulations in the following subsection are carried out on a workstation with a 3.2 GHz CPU and 256 G memory. In addition, all the codes are vectorized in Matlab. To make a fair comparison, only a single thread without any parallel computation is used to carry out the simulations.

\subsection{A Coated Dielectric Cuboid}
\begin{figure}[H]
	\centering
	\includegraphics[width=0.25\textwidth]{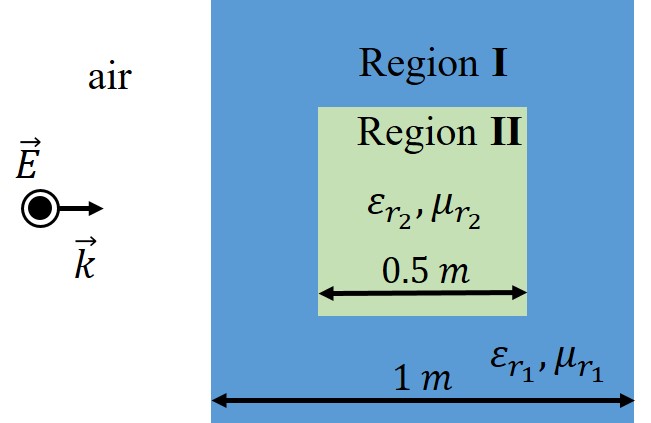}
	\caption{Geometric configurations of the coated dielectric cuboid and its constant parameters.}
	\label{S1}
\end{figure}

In this subsection, { an infinitely long dielectric coated cylinder with square cross section} is considered in Fig. \ref{S1}. The side length of the inner and outer cross section is $0.5$ m and $1$ m, respectively. A TM polarized plane wave with $300$ MHz incidents from the $x$-axis. 

\begin{figure}[H]
	\centering
	\includegraphics[width=0.37\textwidth]{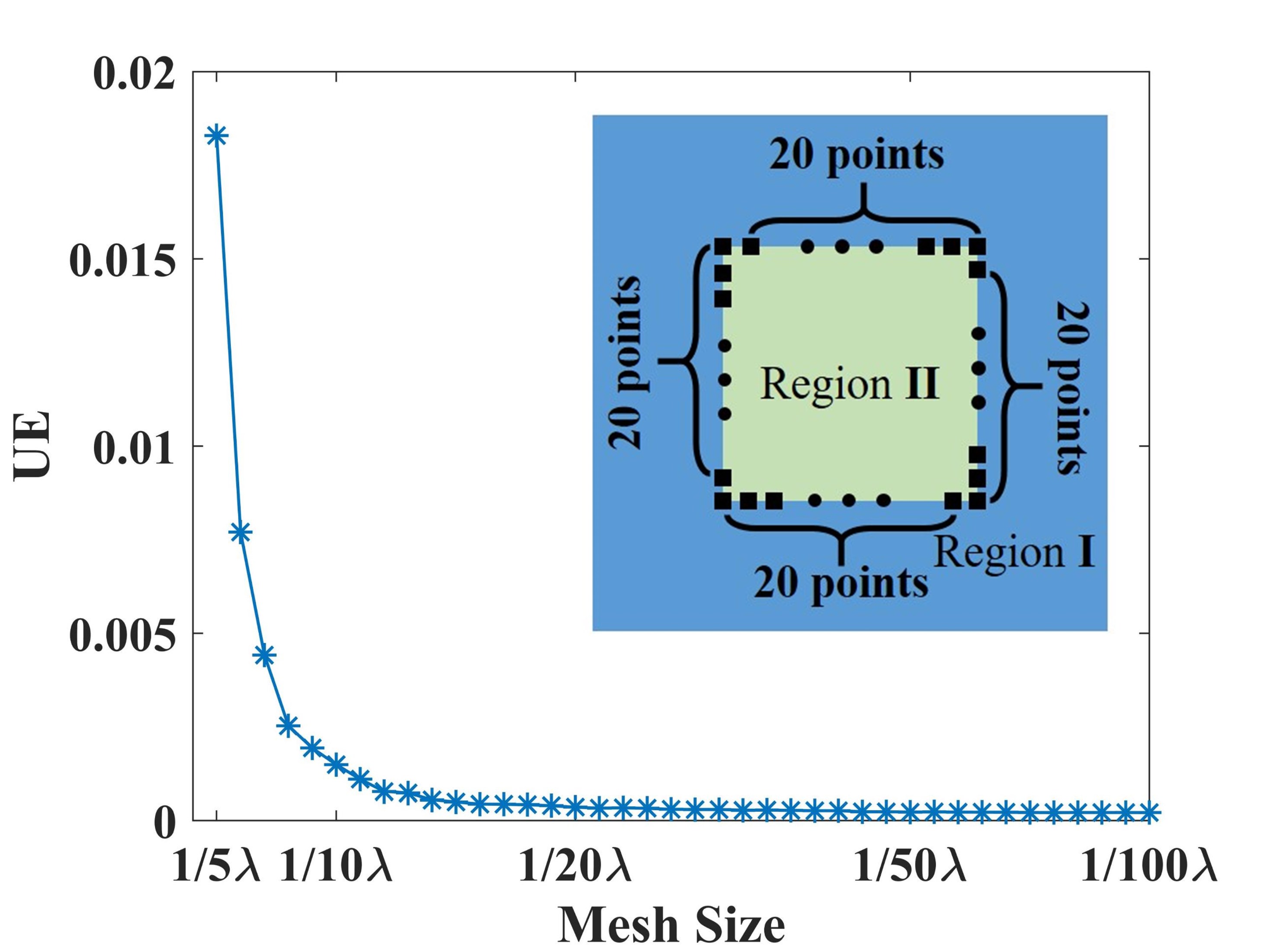}
	\caption{The UE when different mesh is used. $\lambda$ refers to the wavelength in the corresponding region.}
	\label{oversampling}
\end{figure}

{ First, we fixed the parameters in Region I and II as $\varepsilon_{r_1} = 4$, $\mu_{r_1} = 1$, and $\varepsilon_{r_2} = 25$, $\mu_{r_2} = 1$ to verify the numerical stability of the proposed SS-SIE formulation. The uniform error (UE) of the electric fields on the shared boundary is investigated compared with those from the COMSOL when different mesh sizes are used, where the UE is defined as }
\begin{equation}
	{\rm{UE}} = \sqrt {\frac{{{{\sum {\left( {{{\rm{T}}^{{\rm{cal}}}} - {{\rm{T}}^{{\rm{ref}}}}} \right)} }^2}}}{{\sum {{{\left( {{{\rm{T}}^{{\rm{ref}}}}} \right)}^2}} }}} .
\end{equation}
{ We evenly selected 80 sampling points on the boundary (see Fig. \ref{oversampling}), and $\lambda$ in Fig. \ref{oversampling} refers to the wave length in the corresponding region. It is obvious the meshes are nonconformal. It can be found that the UE decreases as the mesh is refined.}

\begin{figure}[H]
	\begin{minipage}[h]{0.495\linewidth}
		\centerline{\includegraphics[scale=0.047]{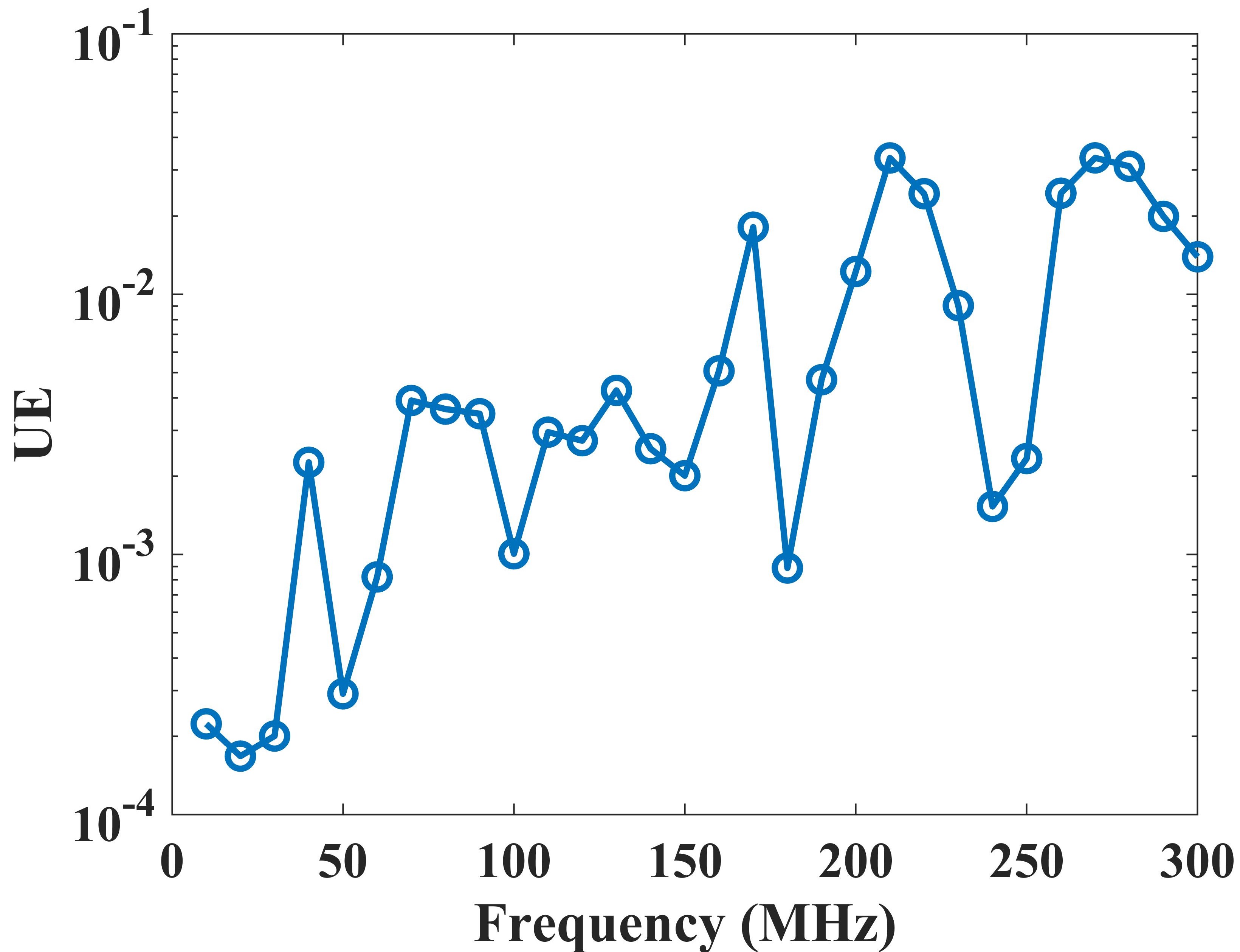}}
		\centerline{(a)}
	\end{minipage}
	\hfill
	\begin{minipage}[h]{0.495\linewidth}
		\centerline{\includegraphics[scale=0.054]{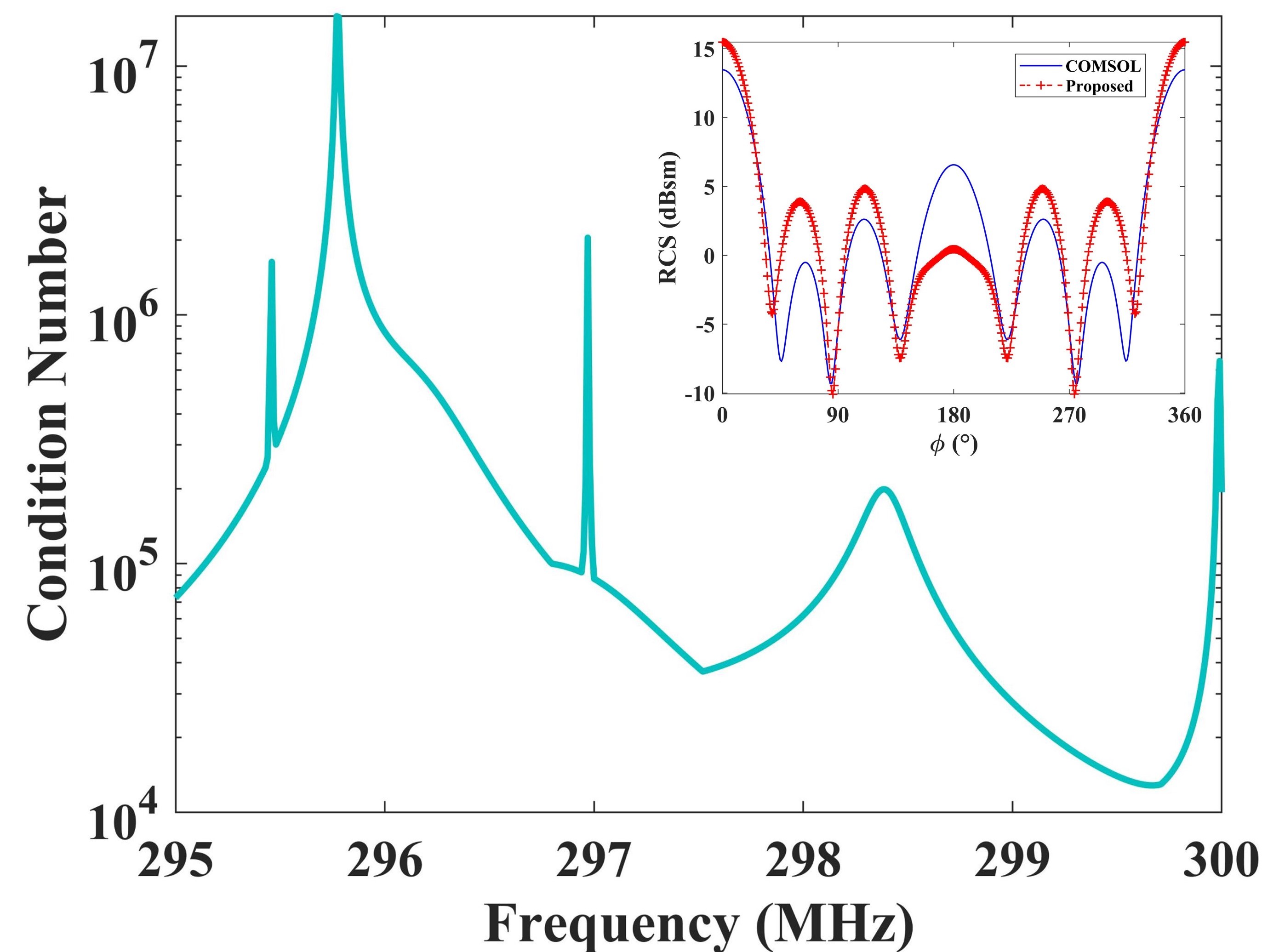}}
		\centerline{(b)}
	\end{minipage}
	\caption{(a) The UE of the proposed SS-SIE formulation from 10 MHz to 300 MHz, where $\varepsilon_{r_1} = 4$, $\mu_{r_1} = 1$, $\varepsilon_{r_2} = 100$, $\mu_{r_2} = 1$, and  the mesh size is $\lambda/10$. (b) The resonant frequency is around 295.77 MHz, where the condition number is much larger.}
	\label{FreR}
\end{figure}

{ The accuracy of the proposed SS-SIE formulation is investigated over a wide frequency band. Fig. \ref{FreR}(a) shows the UE of the sampling points on the shared boundary when the frequency changes from 10 MHz to 300 MHz at the interval of 10 MHz. The parameters of the structure are $\varepsilon_{r_1} = 4$, $\mu_{r_1} = 1$, $\varepsilon_{r_2} = 100$, and $\mu_{r_2} = 1$, respectively. It can be found that the UE is less than 0.06 over a wide frequency band.} { We decreased the interval to 0.01 MHz to sweep frequency response. As shown in Fig. \ref{FreR}(b), the condition number is much larger at the frequency of 295.77 MHz, which indicates the proposed SS-SIE formulation with only the EFIE suffers from the resonance. }

\begin{figure}[H]
	\centering
\includegraphics[width=0.37\textwidth]{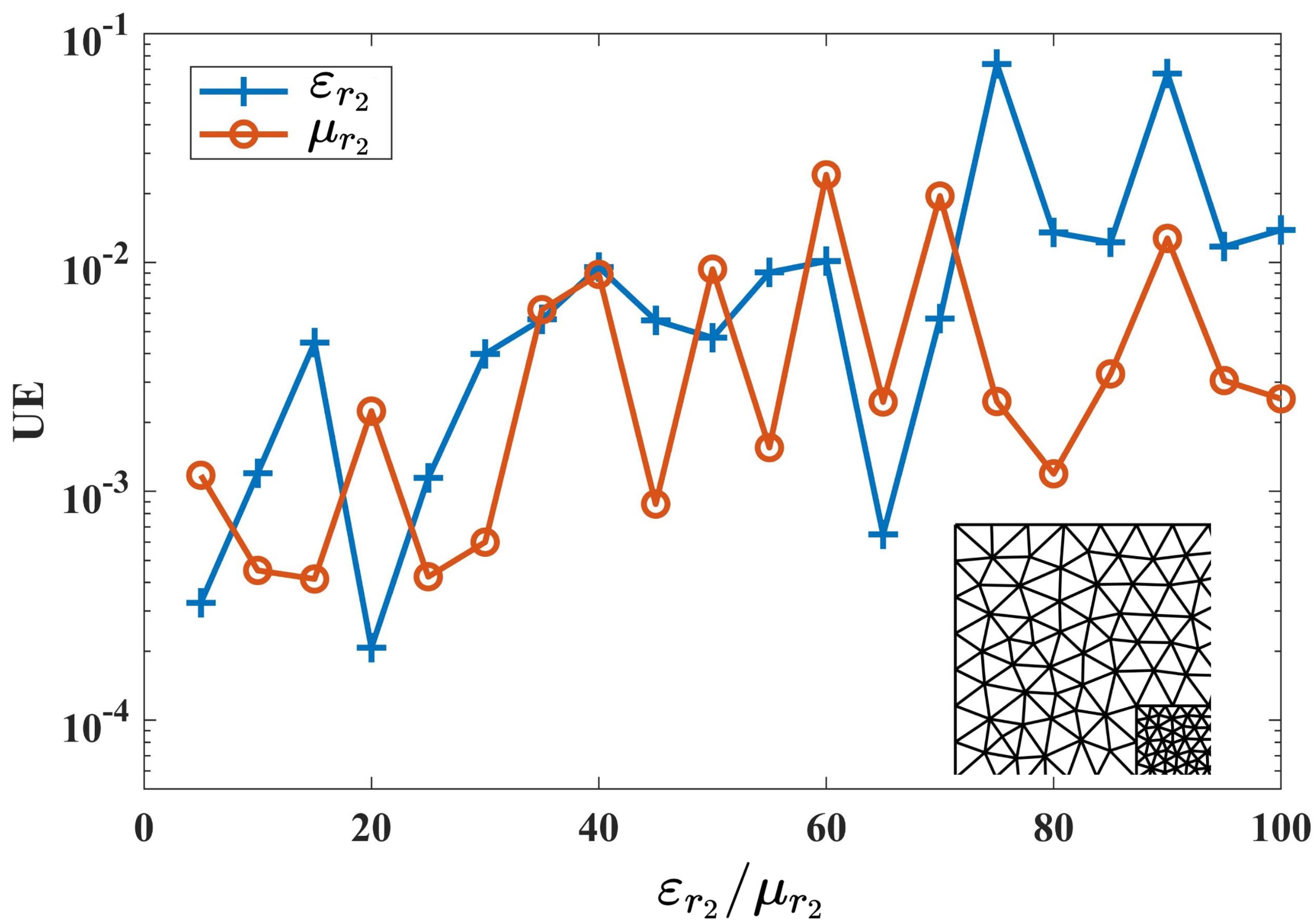}
\caption{The UE when large contrast of dielectric/permeability is used in Region I and Region II. The partial nonconformal meshes are presented when $\varepsilon_{r_2} = 25$ and $\varepsilon_{r_1} = 4$.}
\label{UE}
\end{figure}

{ In addition, we investigated the performance of the proposed SS-SIE formulation with large contrast in terms of permittivity or permeability. We fixed the relative permittivity and permeability in Region I as $\varepsilon_{r_1} = 4$, $\mu_{r_1} = 1$. The relative permittivity in Region II changes from $5$ to $100$ when $\mu_{r_2} = 1$. ${\lambda}/{10}$ is used as the mesh size, where $\lambda$ is the wavelength in the corresponding region. Therefore, the meshes are nonconformal on the shared boundary of the two regions. The UE of the proposed SS-SIE formulation compared with those from the COMSOL is calculated.}
{ Similarly, we swept the relative permeability in Region II from $5$ to $100$ when $\varepsilon_{r_1} = 1$, $\mu_{r_1} = 4$ in Region I, and $\varepsilon_{r_2} = 1$ in Region II. The results are shown in Fig. \ref{UE}, and it can be found that the UE of the near fields is less than $0.1$ when the permittivity or permeability is swept.}

\begin{figure*}
	\begin{minipage}[h]{0.25\linewidth}
		\centerline{\includegraphics[scale=0.06]{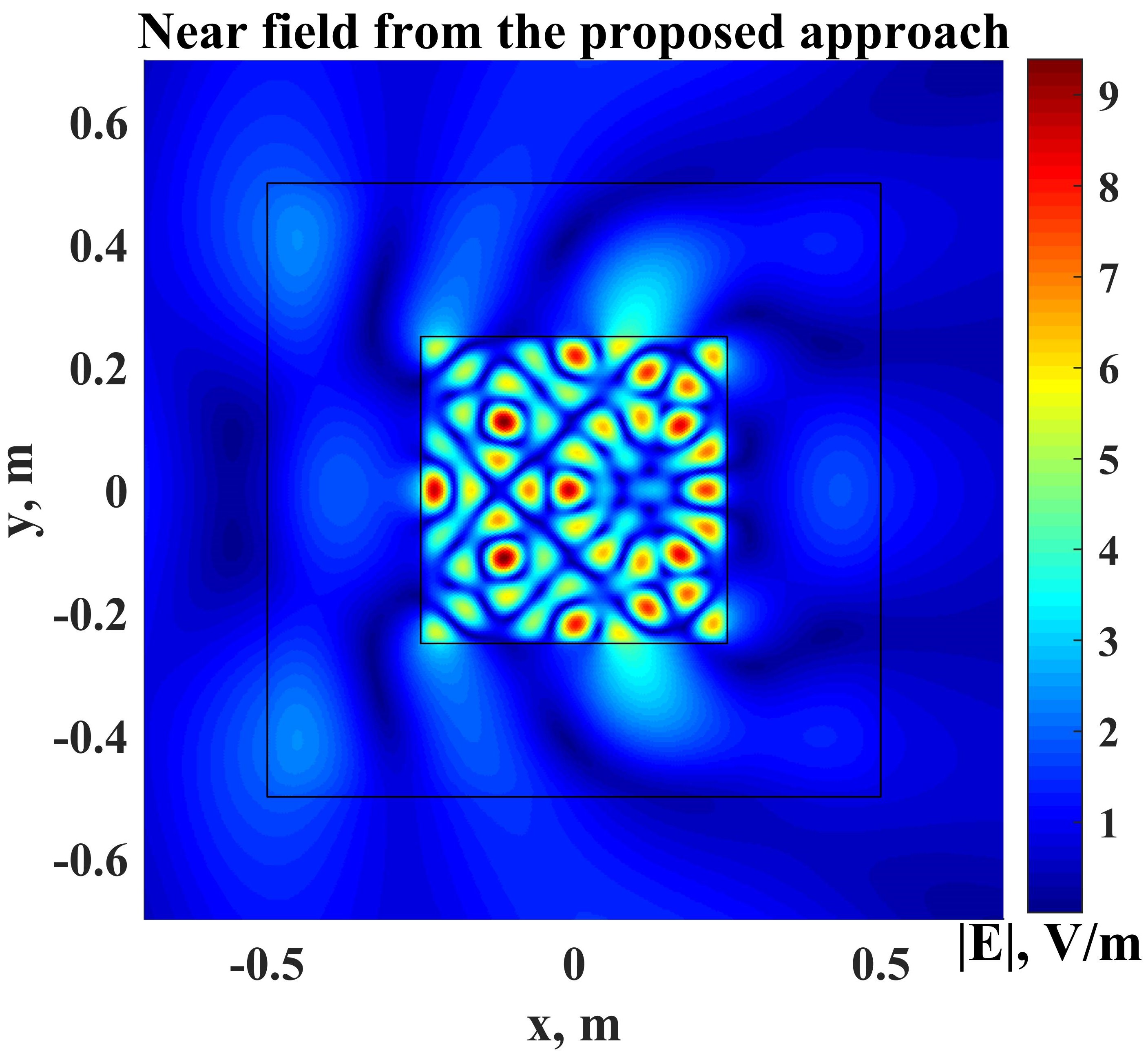}}
		\centerline{(a)}
	\end{minipage}
	\hfill
	\begin{minipage}[h]{0.25\linewidth}
		\centerline{\includegraphics[scale=0.06]{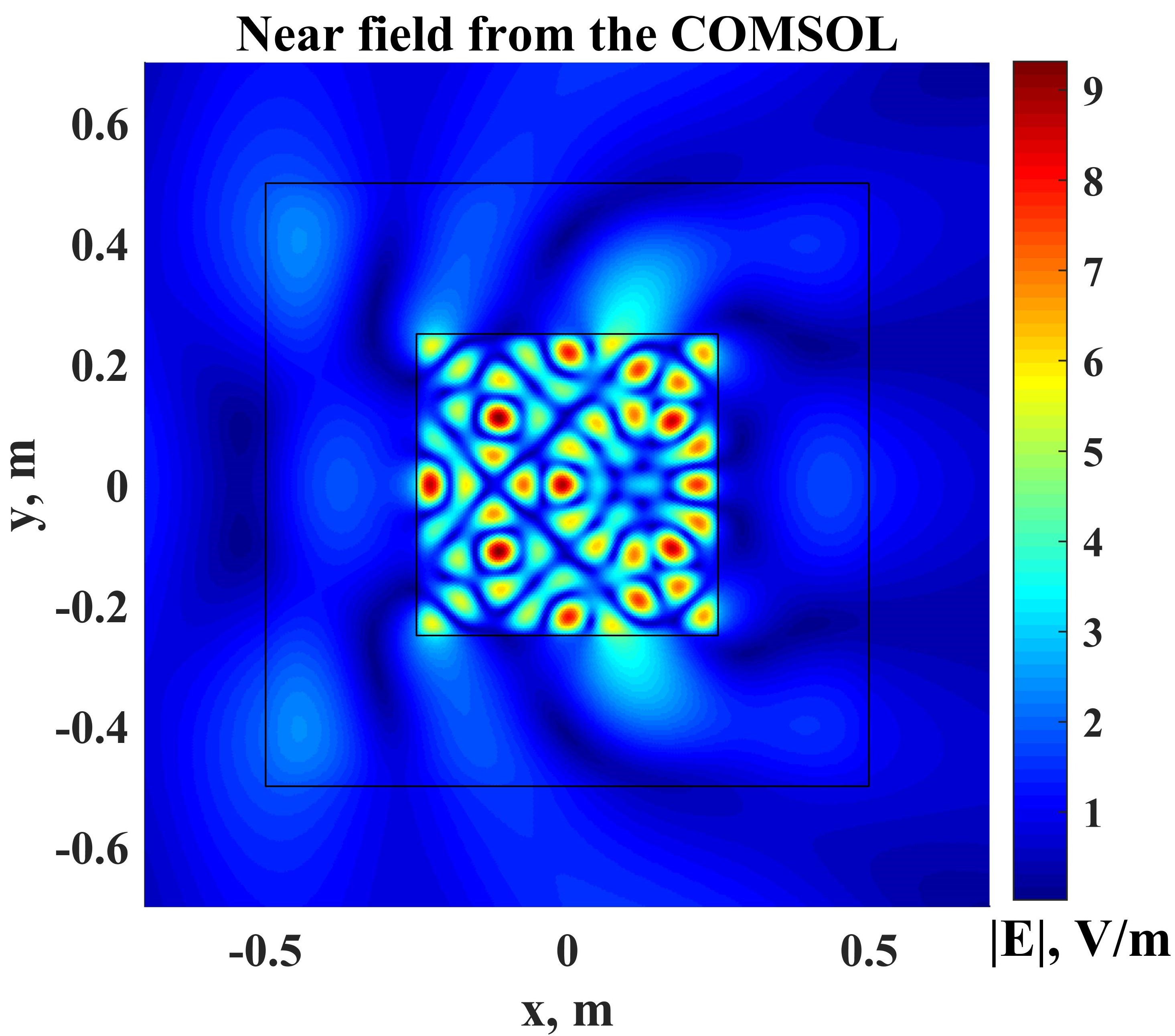}}
		\centerline{(b)}
	\end{minipage}
	\hfill
	\begin{minipage}[h]{0.25\linewidth}
		\centerline{\includegraphics[scale=0.06]{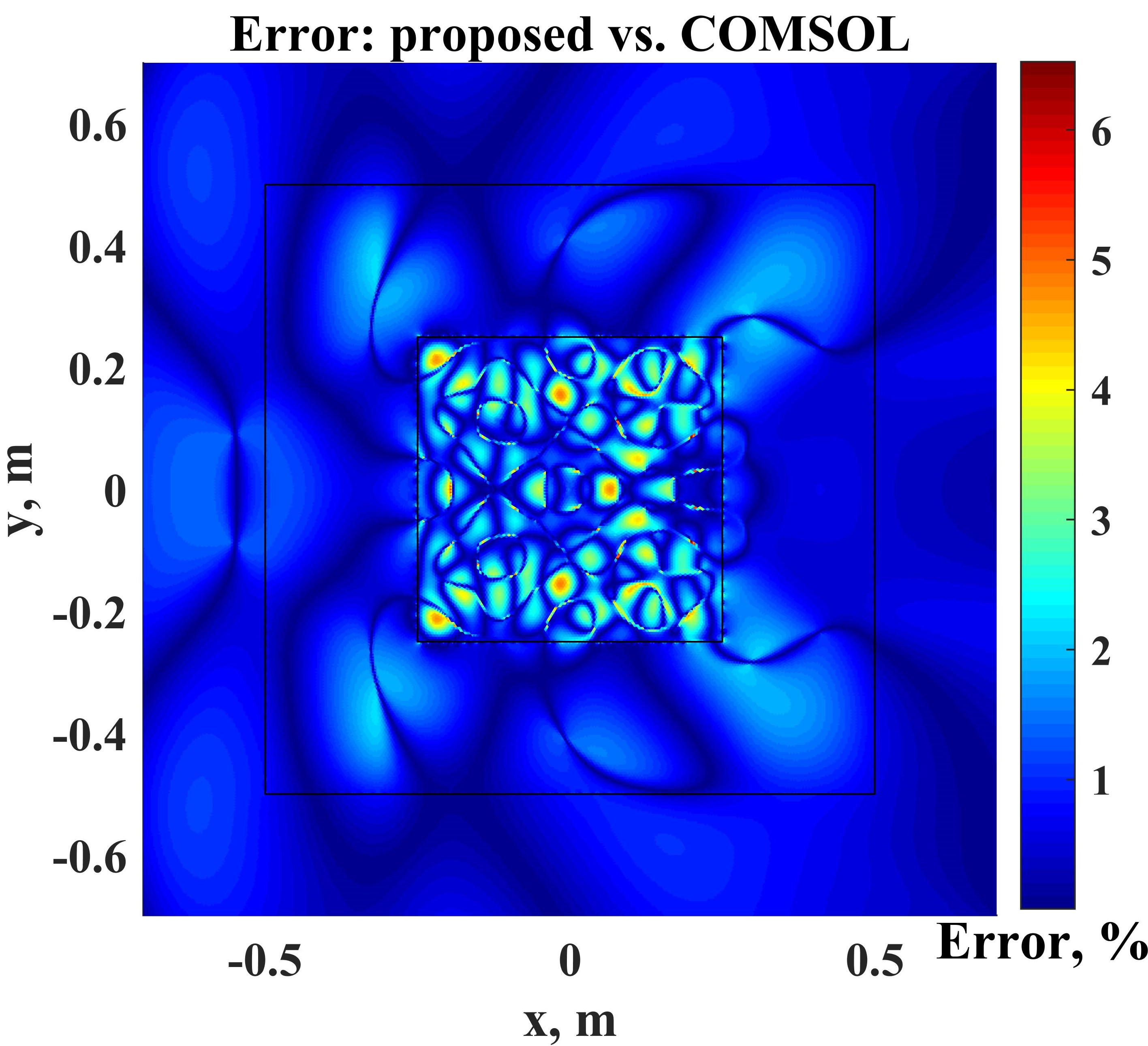}}
		\centerline{(c)}
	\end{minipage} 
	\caption{Near fields obtained from (a) the proposed SS-SIE formulation and (b) the COMSOL, (c) the relative error of near fields obtained from the proposed SS-SIE formulation.}
	\label{N1}
\end{figure*}

{ We further considered the scenario with both magnetic and electric contrast. The parameters in Region I are fixed as $\varepsilon_{r_1} = 2$, $\mu_{r_1} = 2$, and $\varepsilon_{r_1} = 10$, $\mu_{r_1} = 10$ in Region II. Fig. \ref{N1}(a) and (b) show the magnitude of electric fields near the dielectric cuboids obtained from the proposed SS-SIE formulation and the COMSOL. It can be found that field patterns from the two approaches are almost exactly the same and show excellent agreement with each other.}

To quantitatively measure the error, we calculated the relative error of near fields, which is defined as 
\begin{equation}{\label{RE}}
\text{RE} = \frac{\left|\text{E}^{\text {cal }}-\text{E}^{\text {ref }}\right|} {\ {\text {max }}|\text{E}^\text{ref}|},
\end{equation}
where $\text{E}^{\text {ref}}$ are the reference solutions obtained from the COMSOL, $\text{E}^{\text {cal}}$ are the solutions obtained from the proposed SS-SIE formulation, and ${\text {max}}|\text{E}^\text{ref}|$ is the maximum magnitude of the reference solutions, which can guarantee the relative error well defined in the whole computational domain.
As shown in Fig. \ref{N1}(c), the relative error is less than 2$\%$ in the most of computational domain.

\subsection{Domain Decomposition for Large Objects}
\begin{figure}
	\centering
	\includegraphics[width=0.49\textwidth]{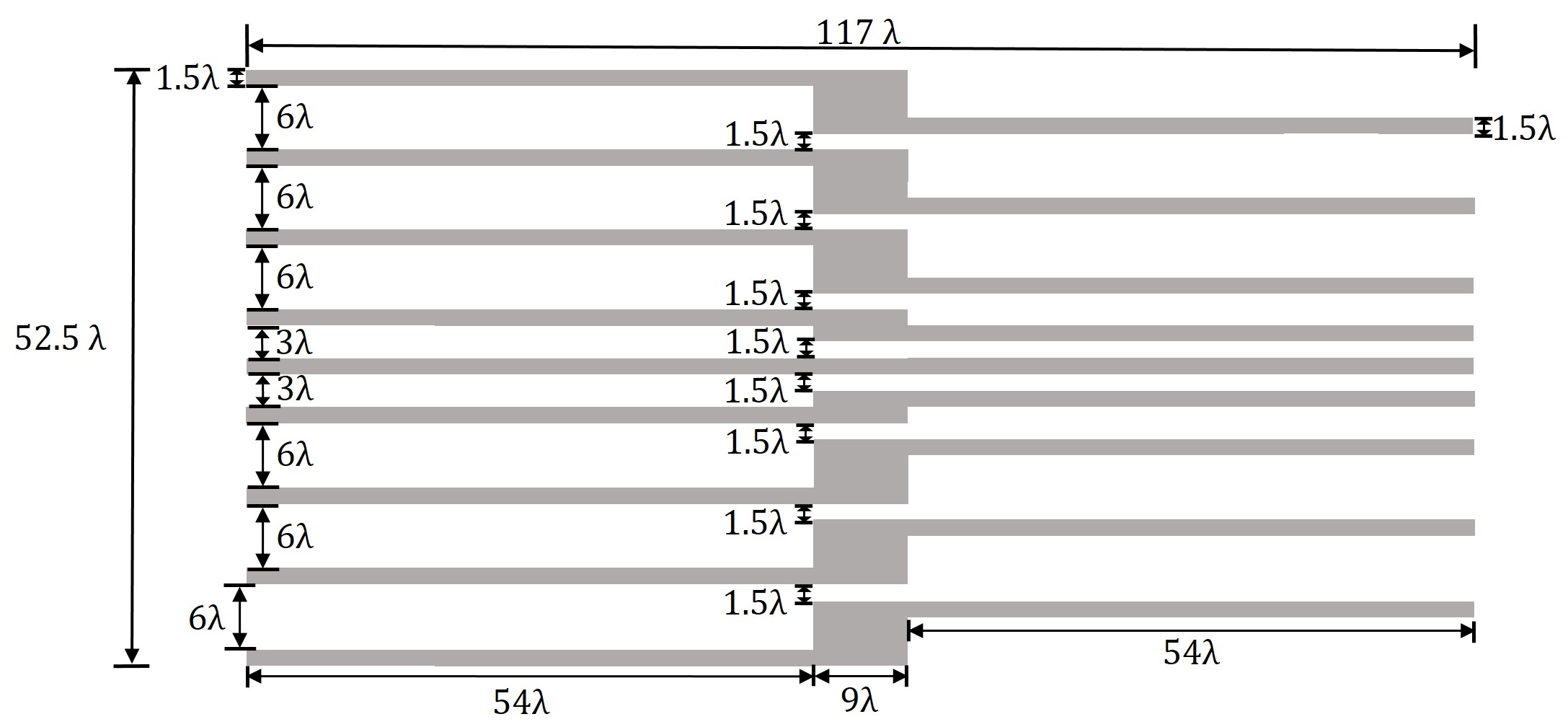}
	\caption{Geometric configurations of nine large lossy objects.}
	\label{S2_O}
\end{figure}
\begin{figure}
	\centering
	\includegraphics[width=0.43\textwidth]{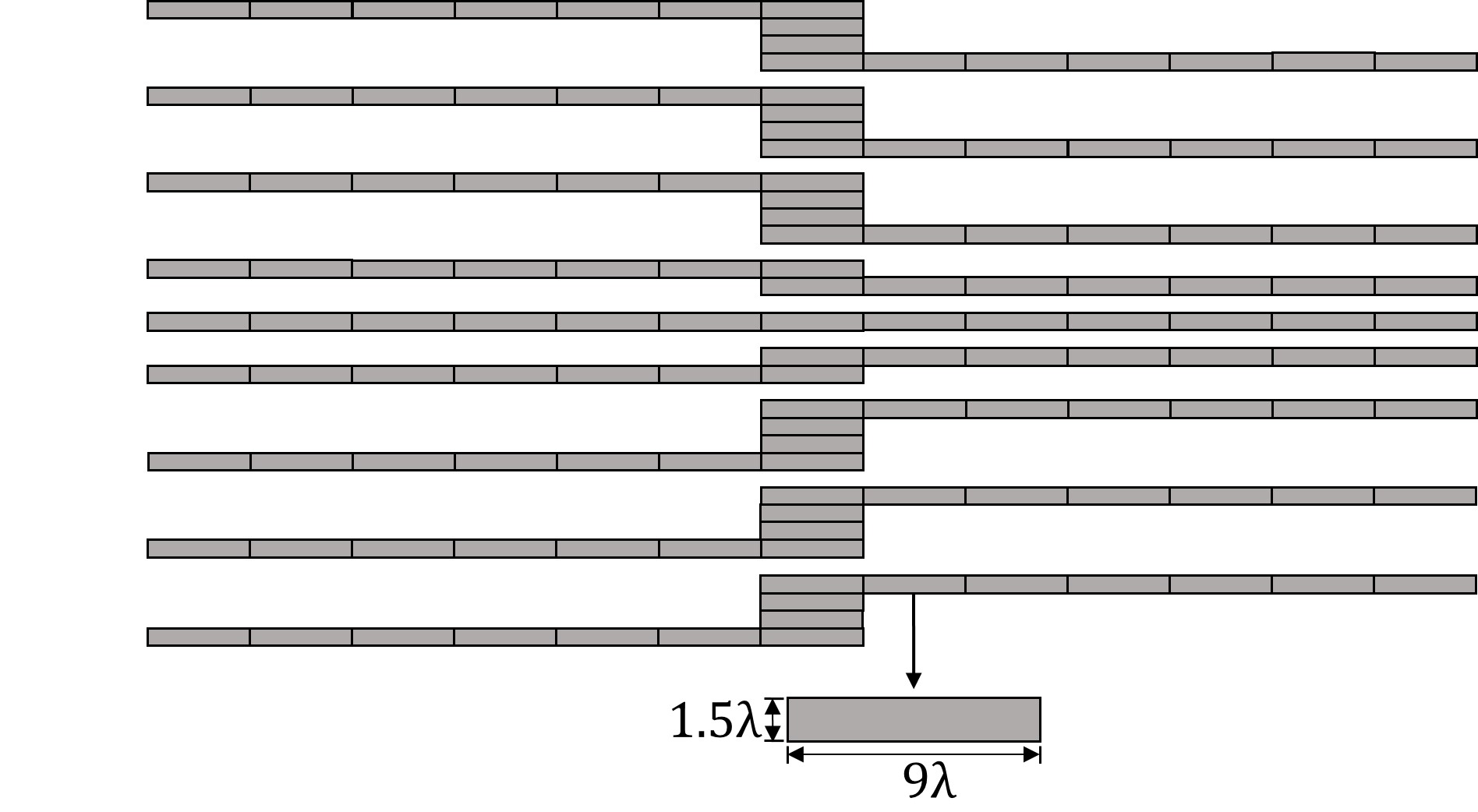}
	\caption{The nine large lossy objects are decomposed into 137 identical small units. The dimension of each unit is 1.5$\lambda$$\times$9$\lambda$.}
	\label{S2_S}
\end{figure}

\begin{figure}
	\centering
	\includegraphics[width=0.45\textwidth]{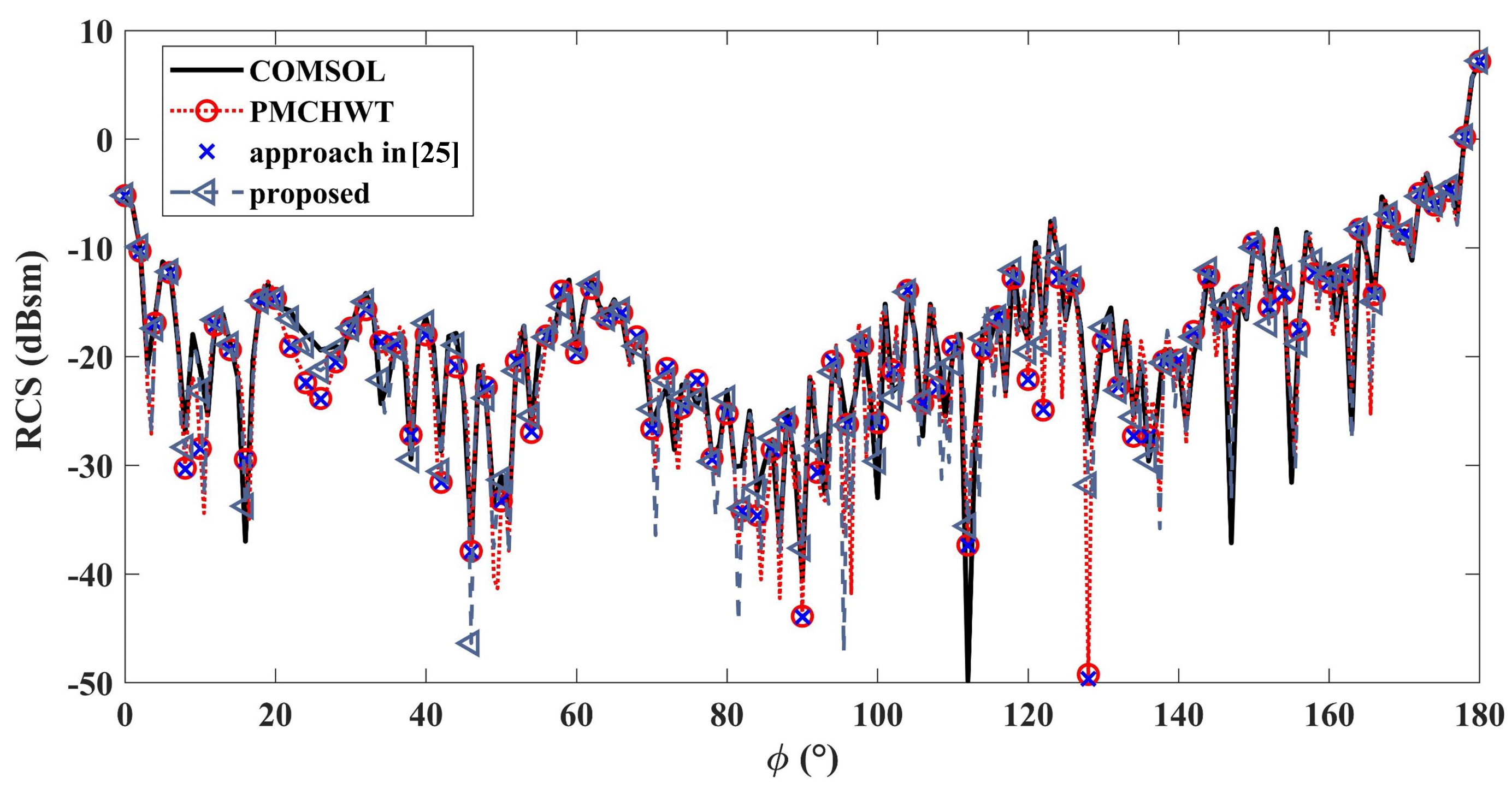}
	\caption{RCS obtained from the COMSOL, the PMCHWT formulation, the SS-SIE formulation in \cite{Zhou2021embedded},  and the proposed SS-SIE formulation.}
	\label{R2}
\end{figure}

The SS-SIE formulation incorporated with the DSAO shows significant performance improvements to model structures with multiple repetitions in large arrays [\citen{Patel2018Macromodeling}], [\citen{Zhou2021embedded}] compared with the PMCHWT formulation. This is because the DSAO for each type of units is only calculated once and then it can be reused to assemble the global DSAO. When identical units are separated with no touched surfaces, the proposed SS-SIE formulation becomes exactly the same as those formulations in \cite{Zhou2021embedded}. Therefore, the proposed SS-SIE formulation inherits this merit. Furthermore, another useful merit is that a large structure can be decomposed into many small identical units to enhance the efficiency.

As shown in Fig. \ref{S2_O}, { nine lossy infinitely long cylinders with the cross section of 117$\lambda$ in length and 52.5$\lambda$ in width are considered}, where $\lambda$ is the wave length in the objects. The relative permittivity is $\varepsilon_r = 9$ and the conductivity is $0.1$ S/m for those objects. A TM polarized plane wave with a frequency of 90 GHz incidents from the $x$-axis.

For the SS-SIE formulation in \cite{Zhou2021embedded}, each object is first replaced by the background medium and an equivalent current density is constructed through inverting a large square matrix $\mathbf{P}$, which is computationally intensive and leads to efficiency degeneration. However, the objects are decomposed into 137 identical small units in the proposed SS-SIE formulation in Fig. \ref{S2_S}. Each unit is $1.5\lambda$ in width and $9\lambda$ in length. Since only one type of units exists in the example, only one DSAO is calculated and then is reused to construct the equivalent model for the large objects. Therefore, it can significantly accelerate the construction of the DSAO. The averaged mesh size used to discretize all boundaries is about $\lambda_0 / 30$, which corresponds to 10 sample segments per wavelength in the lossy dielectric objects.

Fig. \ref{R2} shows the RCS calculated by the COMSOL, the PMCHWT formulation, the SS-SIE formulation in \cite{Zhou2021embedded}, and the proposed SS-SIE formulation. It can be found that the results obtained from the four approaches are in good agreement in most regions. We can observe some   discrepancies in the far fields between the COMSOL and other three formulations, such as around $22^{\circ}$, and $99^{\circ}$. However, all the three integral-equation-based formulations agree well with each other.

\begin{figure*}
	\begin{minipage}[h]{0.3\linewidth}
		\centerline{\includegraphics[scale=0.06]{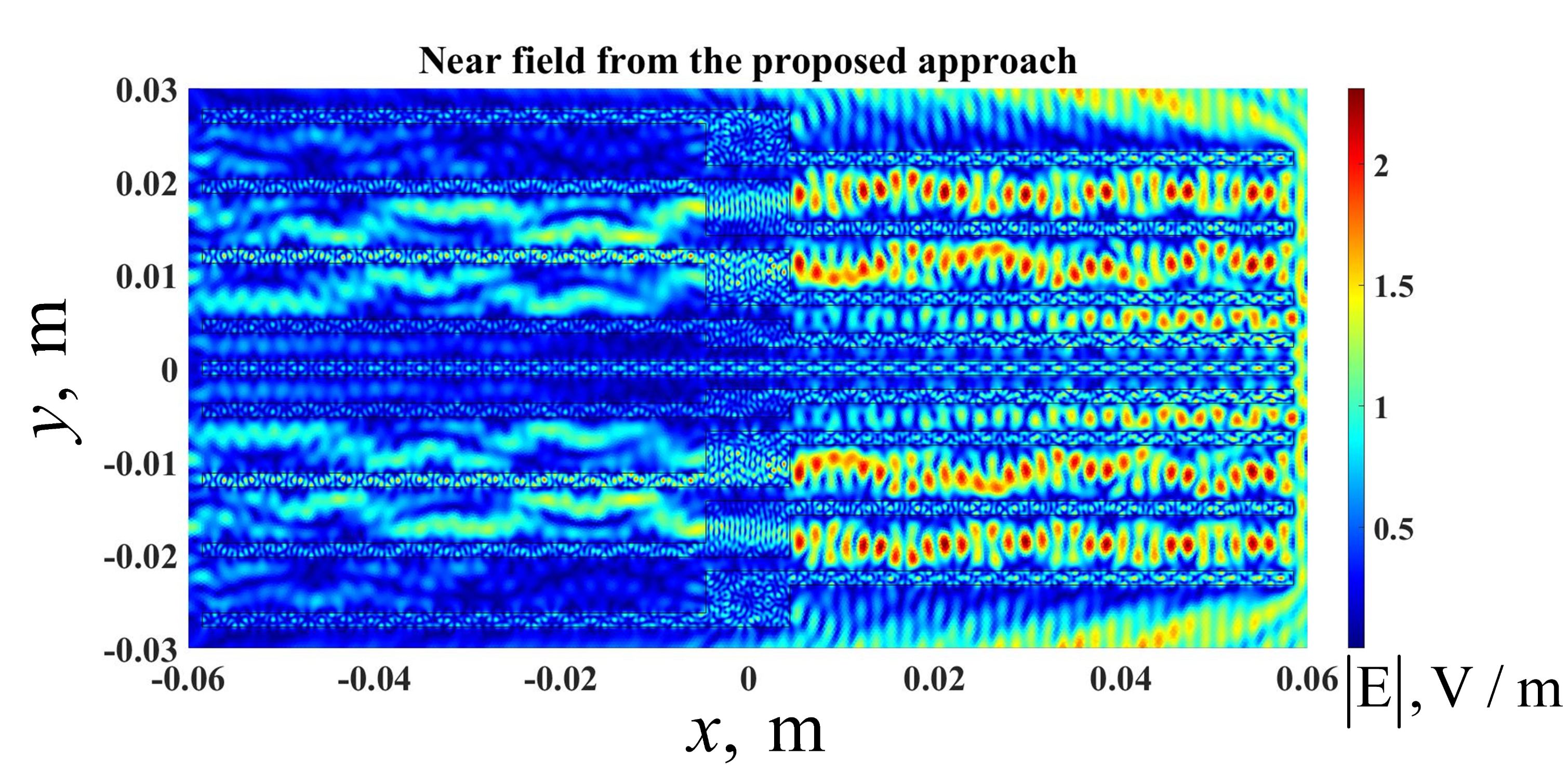}}
		\centerline{(a)}
	\end{minipage}
	\hfill
	\begin{minipage}[h]{0.3\linewidth}
		\centerline{\includegraphics[scale=0.062]{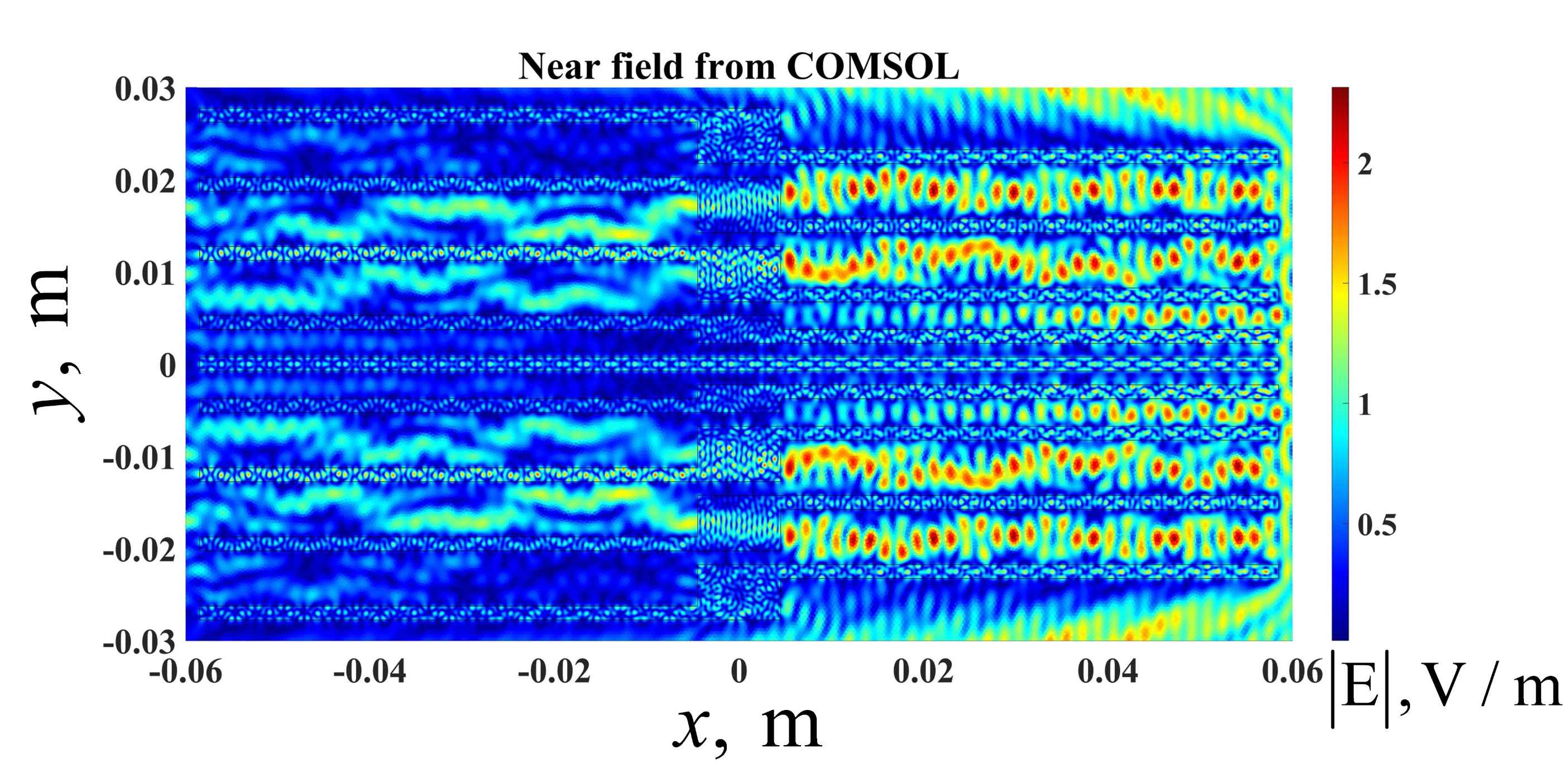}}
		\centerline{(b)}
	\end{minipage}
	\hfill
	\begin{minipage}[h]{0.3\linewidth}
		\centerline{\includegraphics[scale=0.06]{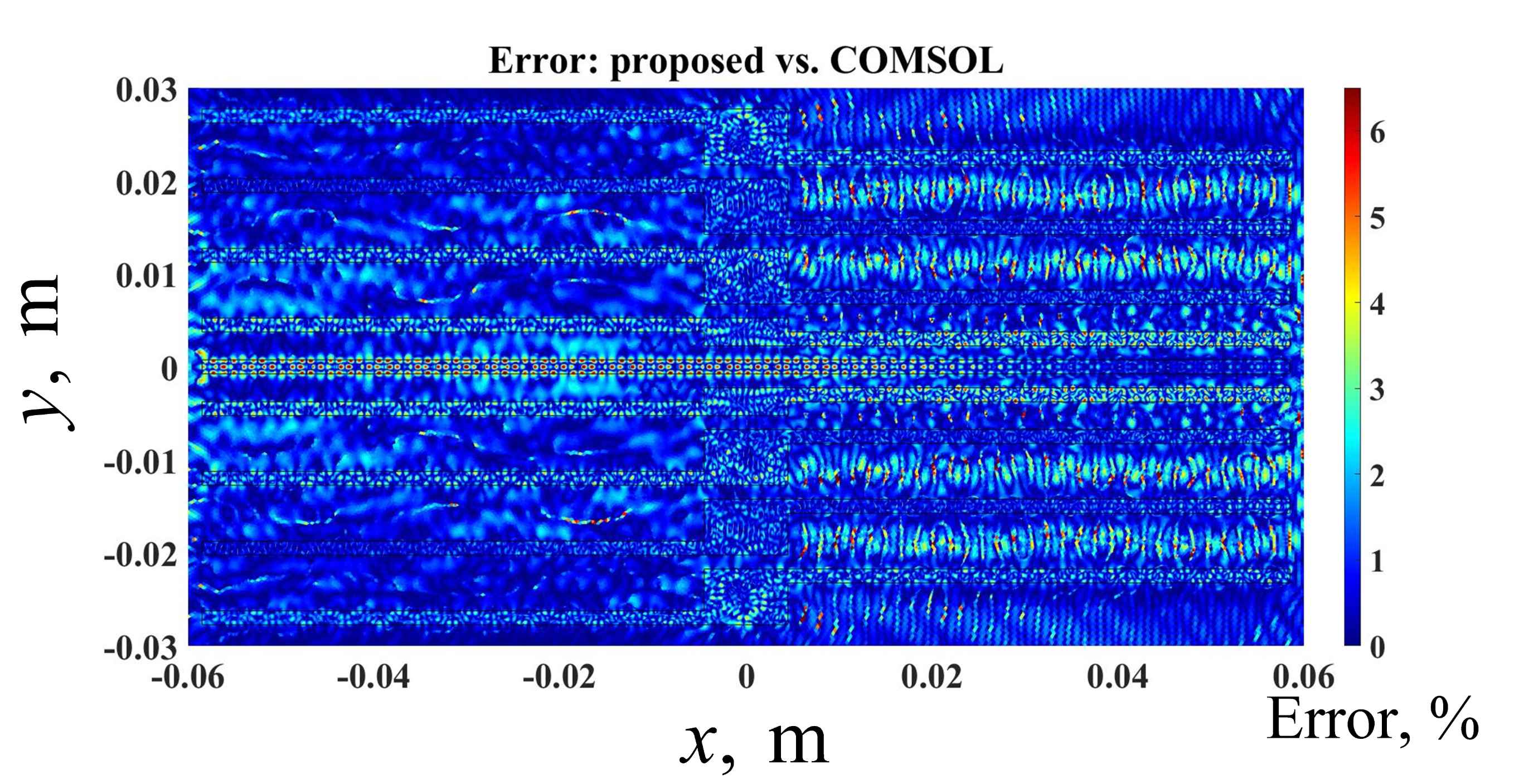}}
		\centerline{(c)}
	\end{minipage} 
	\caption{Near fields obtained from (a) the proposed SS-SIE formulation, (b) the COMSOL, (c) the relative error of near fields obtained from the proposed SS-SIE formulation.}
	\label{N2}
\end{figure*}

Fig. \ref{N2}(a) and (b) show the near fields obtained from the proposed SS-SIE formulation and the COMSOL. It is easy to find that fields pattern obtained from the two approaches are almost the same. In addition, we calculated the relative error of near fields obtained from the proposed SS-SIE formulation in Fig. \ref{N2}(c). In most regions, the relative error is less than 5$\%$ and slightly large in a few points, but only around 7$\%$. Therefore, the proposed SS-SIE formulation can accurately calculate both near and far fields from electrically large objects.

Table I shows the comparison of computational consumption among the PMCHWT formulation, the SS-SIE formulation in \cite{Zhou2021embedded} and the proposed SS-SIE formulation. We only calculate time costs without preprocessing and postprocessing time since they are almost the same for the three formulations. In the proposed SS-SIE formulation, the nine large objects are decomposed into 137 small identical units to construct the DSAO, the time for the generation of DSAO is only { 4.8} seconds and much less than that of the SS-SIE formulation in \cite{Zhou2021embedded}, which requires { 5,351.4} seconds. Time costs for matrices filling and matrices solving are { 6,500.2} seconds and { 793.3} seconds for the proposed SS-SIE formulation, respectively. The time are more compared with the formulation in \cite{Zhou2021embedded}, which requires { 5,813.4} seconds and { 569.5} seconds, respectively. The reason is that more unknowns exist on the interfaces between two adjacent units in the proposed SS-SIE formulation. { It should be noted that the matrix solving time included the calculation of the $\mathbf{E}$ unknowns and the reconstruction of $\mathbf{H}$ unknowns through (\ref{Y_Defined}).} However, the overall time cost for the proposed SS-SIE formulation is only 7,303.4 seconds, which shows significant performance improvement in terms of CPU time compared with 11,741.2 seconds for the SS-SIE formulation in \cite{Zhou2021embedded} and 22,988.9 seconds for the PMCHWT formulation. In addition, memory consumption of the proposed SS-SIE formulation, 63,322 MB, is slightly more compared with that of the formulation in \cite{Zhou2021embedded}, 52,751 MB, because more unknowns are needed to be stored. However, in comparison with that of the PMCHWT formulation, 92,297 MB, the proposed SS-SIE formulation still shows significant performance improvement.
\renewcommand\arraystretch{1.2}
\begin{table}[H]
	\centering
	\caption{Comparison of computational cost for the PMCHWT formulation, the SS-SIE formulation in [44], and the proposed SS-SIE formulation }\label{table1}
	\begin{tabular}{c|c|c|c}
		\hline
		&\textbf{PMCHWT}&\textbf{Formulation in \cite{Zhou2021embedded}} &\textbf{Proposed} \cr 
		\hline 
		\hline
		\textbf{Total Time }  [s]             &22,988.9             &11,734.3     &7,298.3     \\
		\hline
		Time for $\mathbb{Y}_s$              &\multirow{2}{*}{-}            &\multirow{2}{*}{5,351.4}            &\multirow{2}{*}{4.8}            \\
		Generation [s] & & &\\
		\hline
		Time for Ma-      &\multirow{2}{*}{21,563.0}           & \multirow{2}{*}{5,813.4}            & \multirow{2}{*}{6,500.2}       \\  
		trices Filling [s]&&&\\
		\hline
		Time for Ma-         &\multirow{2}{*}{1,425.9}            &\multirow{2}{*}{569.5}     &  \multirow{2}{*}{793.3}   \\
		trix Solving [s] &&&\cr
		\hline
		\hline
		\textbf{Memory Con-}                   & \multirow{2}{*}{92,297}             & \multirow{2}{*}{52,751}    &\multirow{2}{*}{63,322}         \\
		\textbf{sumption} [MB]&&&\\
		\hline
		\textbf{Number of}                 & \multirow{2}{*}{60,000}             & \multirow{2}{*}{30,000}    &\multirow{2}{*}{39,456}     \\
		\textbf{Unknowns} &&&\\
		\hline
	\end{tabular}
\end{table}

\subsection{Layered Penetrable Objects}
\begin{figure}[H]
	\centering
	\includegraphics[width=0.48\textwidth]{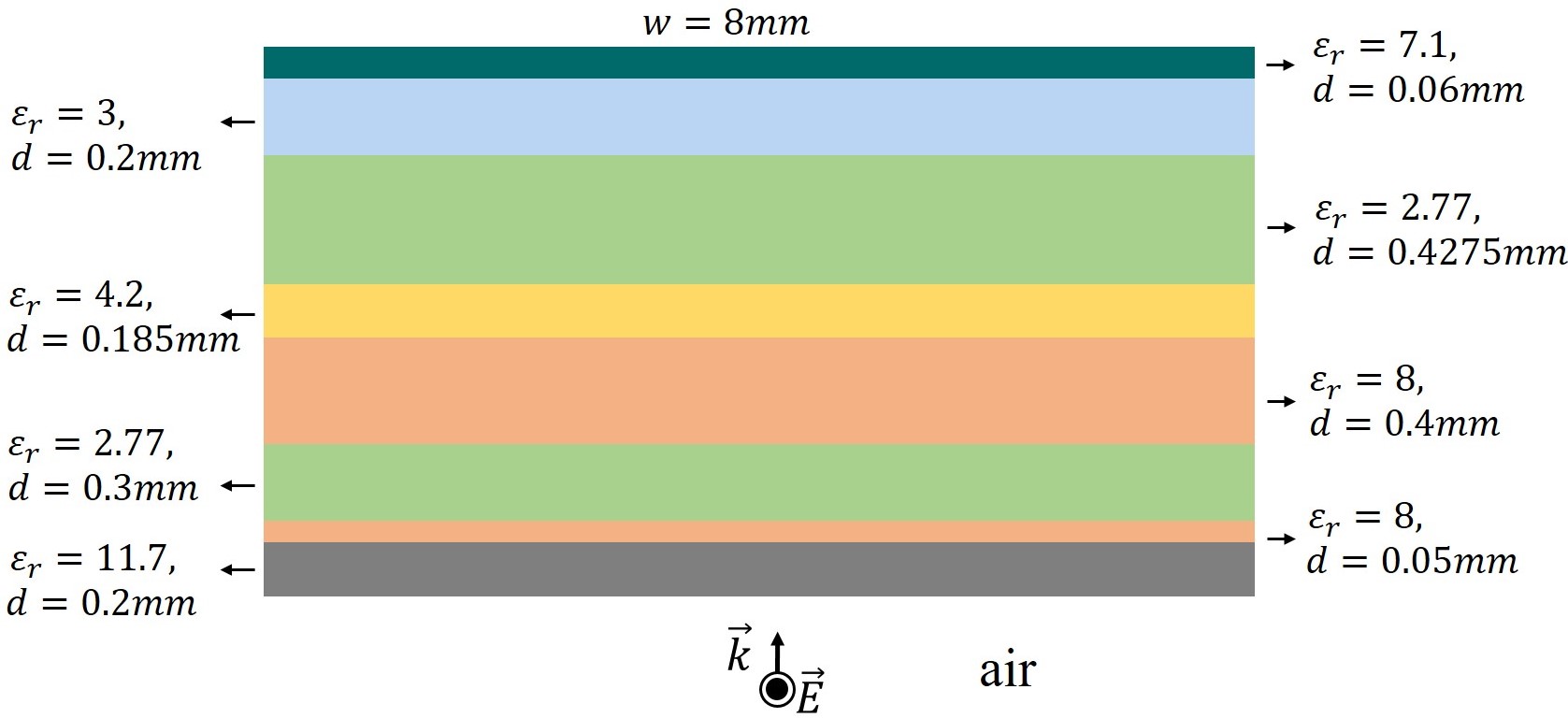}
	\caption{An 8-layered structure including eight dielectric objects with different permittivity as marked in the figure.}
	\label{S3_O}
\end{figure}
\begin{figure}[H]
	\centering
	\includegraphics[width=0.4\textwidth]{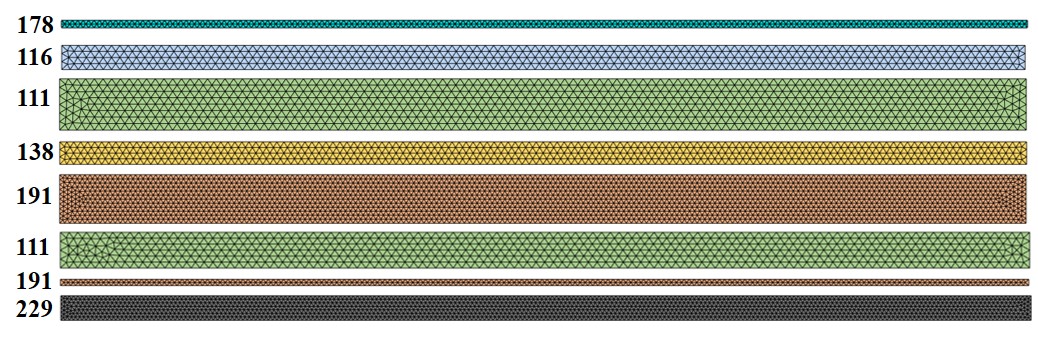}
	\caption{The nonconformal meshes for the layered objects and the numbers on the left of the structure indicate the count of segments on the boundaries  used in our simulations. Note that a small gap is added and the triangle edges are presented to better visualize the nonconformal meshes. Only the boundary segments are used in our simulations.}
	\label{S3_M}
\end{figure}

{ An infinitely long dielectric cylinder with the cross section of 8-layered rectangles} is considered in this subsection. The relative permittivity of each layer is 7.1, 3, 2.77, 4.2, 8, 2.77, 8 and 11.7 from the top to bottom layers in Fig. \ref{S3_O}. The electrical size of this structure is around 6.7$\lambda$ in length and 1.5$\lambda$ in width. A plane wave with the frequency of 250 GHz incidents from the $y$-axis. The mesh size is chosen based on the wave length in the corresponding layer, which is $\lambda_0/27$, $\lambda_0/17$, $\lambda_0/16$, $\lambda_0/20$, $\lambda_0/28$, $\lambda_0/16$, $\lambda_0/28$ and $\lambda_0/34$, respectively. Fig. \ref{S3_M} shows the meshes used in our simulation. A small gap is added and the triangle edges are presented to better visualize the nonconformal meshes. It should be noted that only the boundary segments are used in our simulations. The numbers marked on the LHS indicate the count of segments on the shared boundaries. 

\begin{figure}[H]
	\centering
	\includegraphics[width=0.45\textwidth]{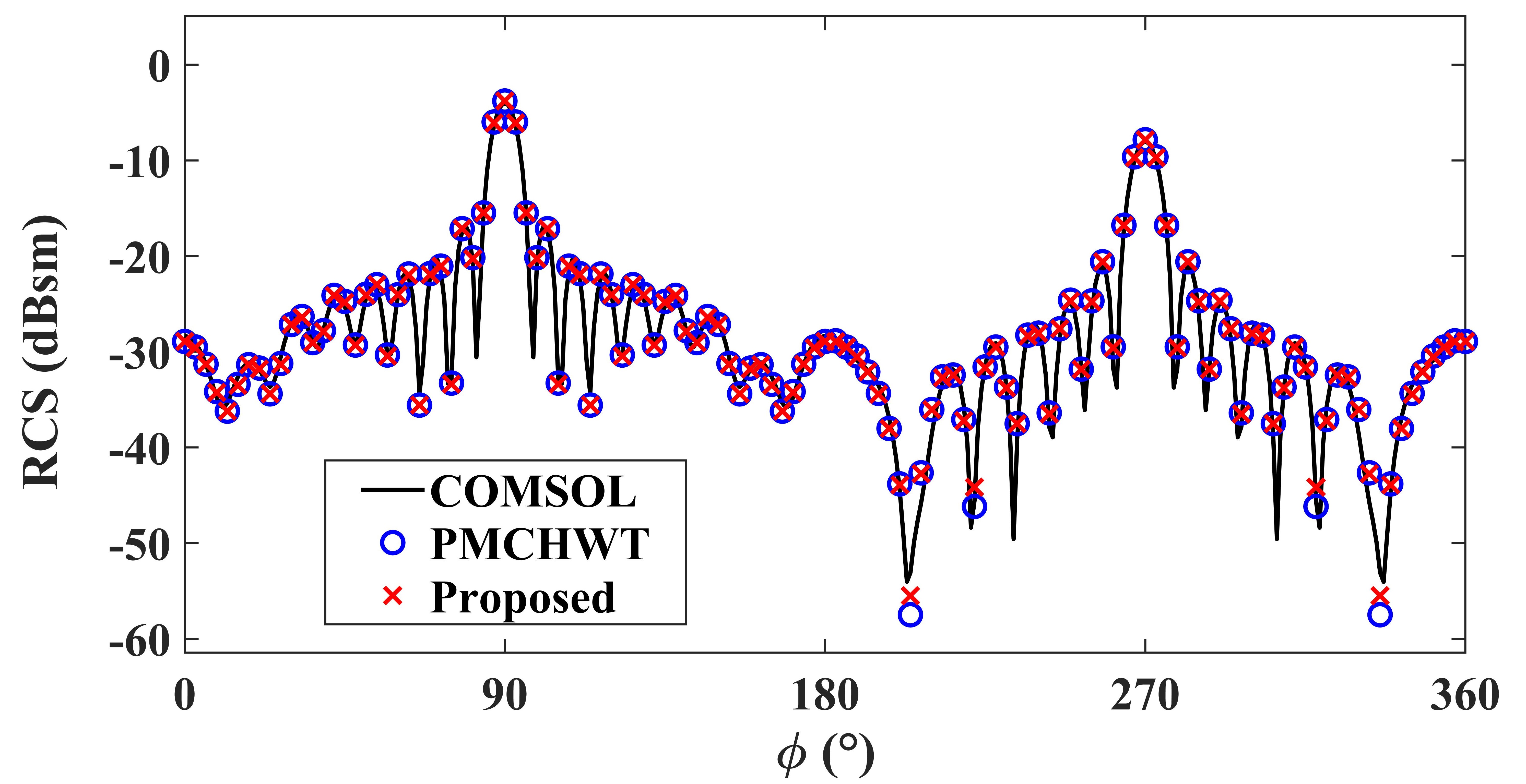}
	\caption{The RCS obtained from the COMSOL, the PMCHWT formulation, and the proposed SS-SIE formulation.}
	\label{R3}
\end{figure}

\renewcommand\arraystretch{1.4}
\begin{table}[H]
	\centering
	\caption{Comparison of computational cost for the PMCHWT formulation and the proposed SS-SIE formulation }\label{table2}
	\begin{tabular}{l|c|c}
		\hline
		&\textbf{PMCHWT}&\textbf{Proposed}  \cr 
		\hline 
		\hline
		\textbf{Total Time }  [s]             &399.7             &123.8          \\
		\hline
		Time for $\mathbb{Y}_s$ Generation [s]              &{-}            &{40.3}                        \\
		\hline
		Time for Matrices Filling [s]     &{390.7}           & {81.9}                   \\  
		\hline
		Time for Matrix Solving [s]         &{8.9}            &{1.6}        \\
		\hline
		\hline
		\textbf{Memory Consumption} [MB]                   & {995}             & {401}             \\
		\hline
		\textbf{Number of Unknowns}                 & {7,572}             & {2,600}         \\
		\hline
	\end{tabular}
\end{table} 
Fig. \ref{R3} shows the RCS obtained from the COMSOL, the PMCHWT formulation, and the proposed SS-SIE formulation. It can be found that numerical results from the three approaches are in good agreement. Therefore, the proposed SS-SIE formulation can obtain accurate far fields for layered composite objects with nonconformal meshes.

\begin{figure*}
	\begin{minipage}[h]{0.32\linewidth}
		\centerline{\includegraphics[scale=0.02]{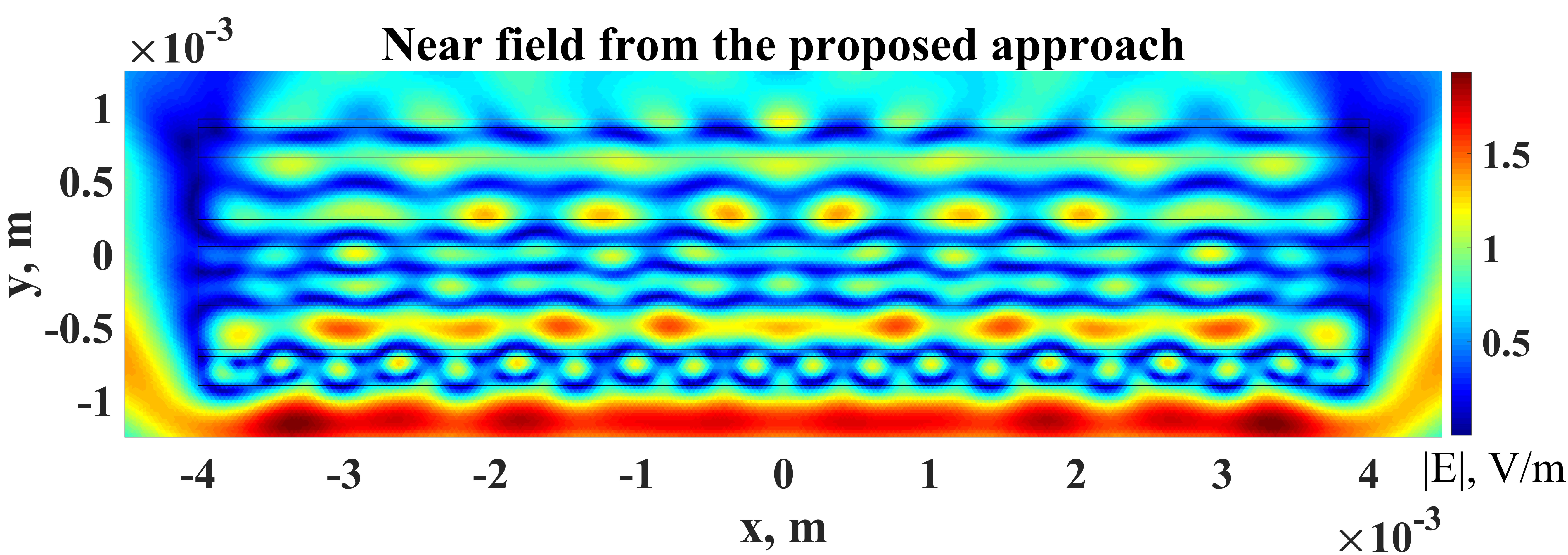}}
		\centerline{(a)}
	\end{minipage}
	\hfill
	\begin{minipage}[h]{0.32\linewidth}
		\centerline{\includegraphics[scale=0.02]{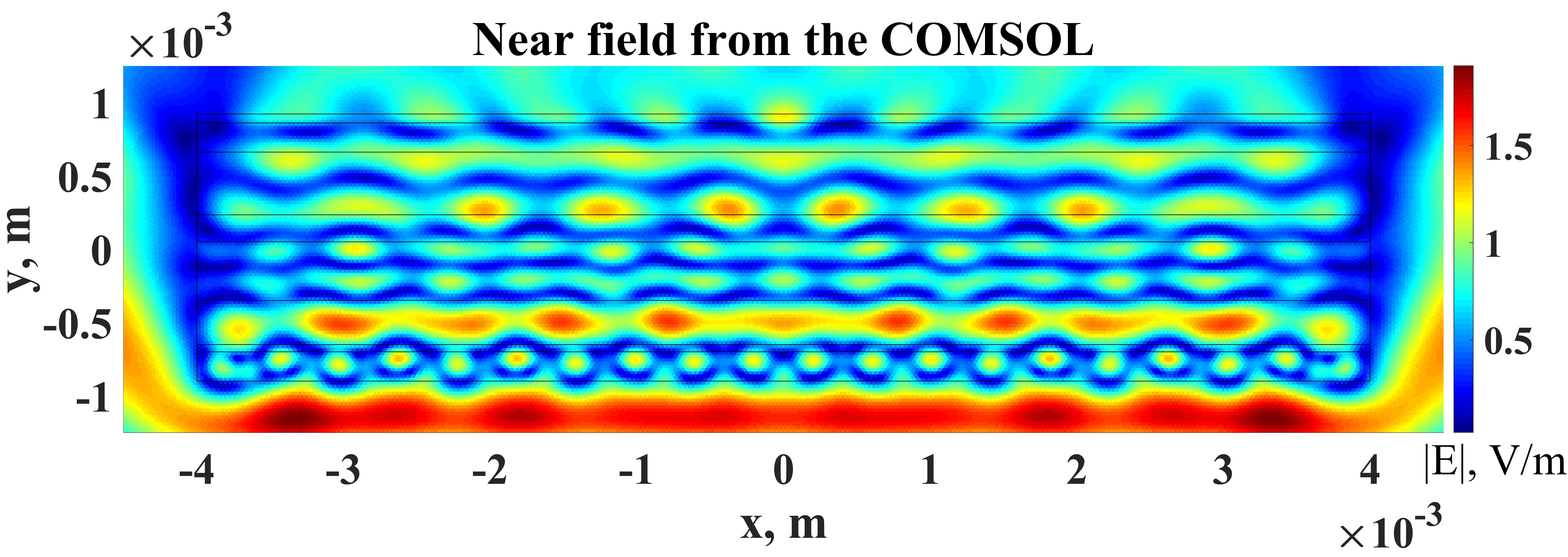}}
		\centerline{(b)}
	\end{minipage}
	\hfill
	\begin{minipage}[h]{0.32\linewidth}
		\centerline{\includegraphics[scale=0.06]{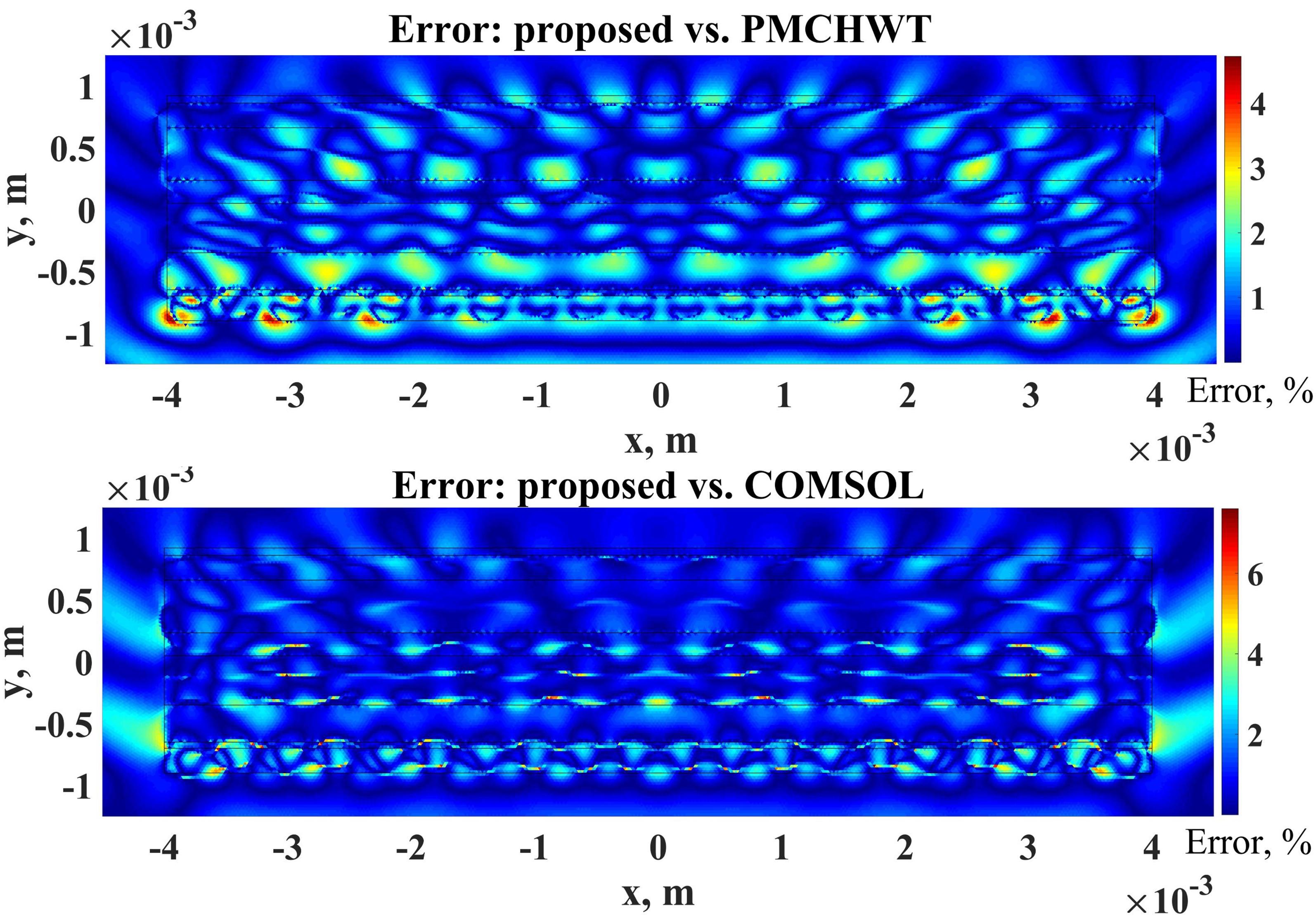}}
		\centerline{(c)}
	\end{minipage} 
	\caption{(a) Near fields obtained from the proposed SS-SIE formulation, (b) near fields obtained from the COMSOL, (c) the relative error of near fields.}
	\label{N3}
\end{figure*}
Fig. \ref{N3} shows near fields obtained from the proposed SS-SIE formulation and the COMSOL. It can be found that near fields obtained from the two formulations show good agreement in Fig. \ref{N3}(a) and (b). We also calculated the relative error of near fields obtained from the proposed SS-SIE formulation compared with those from the COMSOL and the PMCHWT formulation. As shown in Fig. \ref{N3}(c), the relative error in most regions is less than 3$\%$, which shows that the proposed SS-SIE formulation can obtain accurate near fields like other numerical techniques.

As shown in Table II, to finish the simulation, the PMCHWT formulation uses { 399.7} seconds and 7,572 unknowns, where the averaged length of segments is $\lambda_0/34$, and only the conformal meshes can be used. However, the proposed SS-SIE formulation only takes { 123.8} seconds, { where 40.3 seconds are used for $\mathbb{Y}_s$ generation, $81.9$ seconds are for matrix filling, and $1.6$ seconds are for matrix equation solving.} Since the nonconformal meshes are used, and only single electric source exists, the proposed SS-SIE formulation only take $2600$ unknowns. Therefore, it shows significant performance improvement.

\subsection{A Composite Structure with Partially Connected Penetrable and PEC Objects}
\begin{figure}
	\centering
	\includegraphics[width=0.32\textwidth]{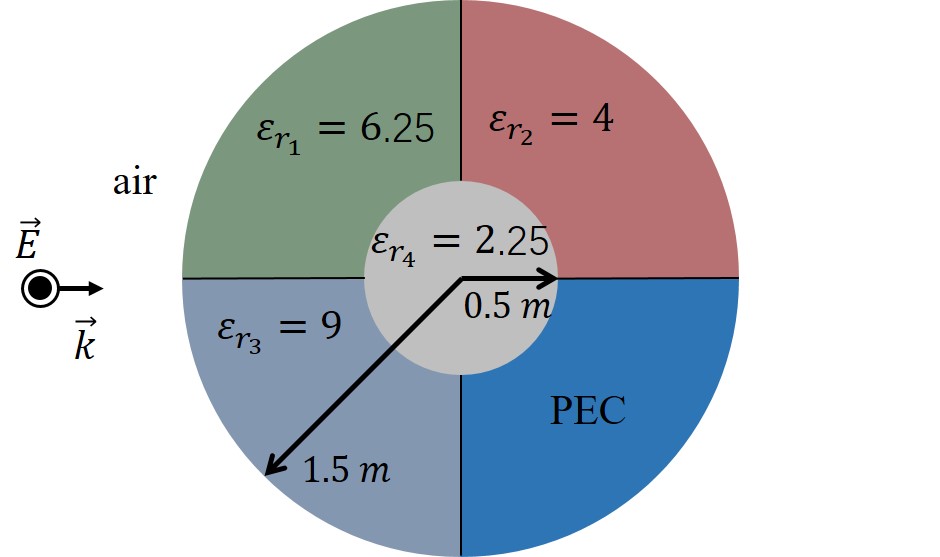}
	\caption{A composite structure consisting of four dielectric objects and one PEC object.}
	\label{S4}
\end{figure}
\begin{figure}
	\centering
	\includegraphics[width=0.21\textwidth]{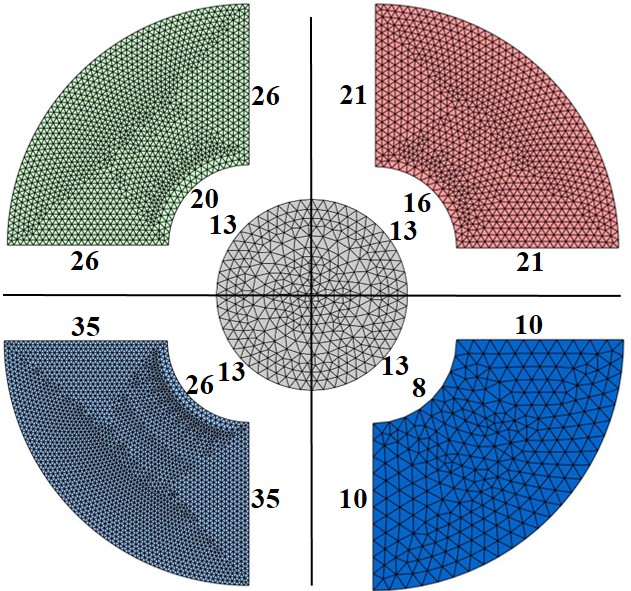}
	\caption{The nonconformal meshes for the composite object and the numbers indicate the count of the segments on the shared boundaries used in our simulations. Note that a small gap is added and the triangle edges are presented to better visualize the nonconformal meshes. Only the boundary segments are used in our simulations.}
	\label{S4_m}
\end{figure}

{ An infinitely long cylinder with the cross section of five different shapes} is considered in Fig. \ref{S4}. There are four dielectric objects with the relative permittivity of $\varepsilon_{r_1} = 6.25$, $\varepsilon_{r_2} = 4$, $\varepsilon_{r_3} = 9$, $\varepsilon_{r_4} = 2.25$ and a PEC object. The radii of inner dielectric object is 0.5 m and 1.5 m for the four outer quarter concentric objects. The averaged length of segments to discretize the five objects is $\lambda_0/25$, $\lambda_0/20$, $\lambda_0/30$, $\lambda_0/15$ and $\lambda_0/10$, respectively. Similar to previous example, a small gap is added and the triangle edges are presented to better demonstrate the nonconformal meshes. The numbers of segments to discretize the corresponding boundary are marked in Fig. \ref{S4_m}, which are selected based on the constant parameters. A plane wave with the frequency of 300 MHz incidents from the $x$-axis.

\begin{figure}
	\centering
	\includegraphics[width=0.45\textwidth]{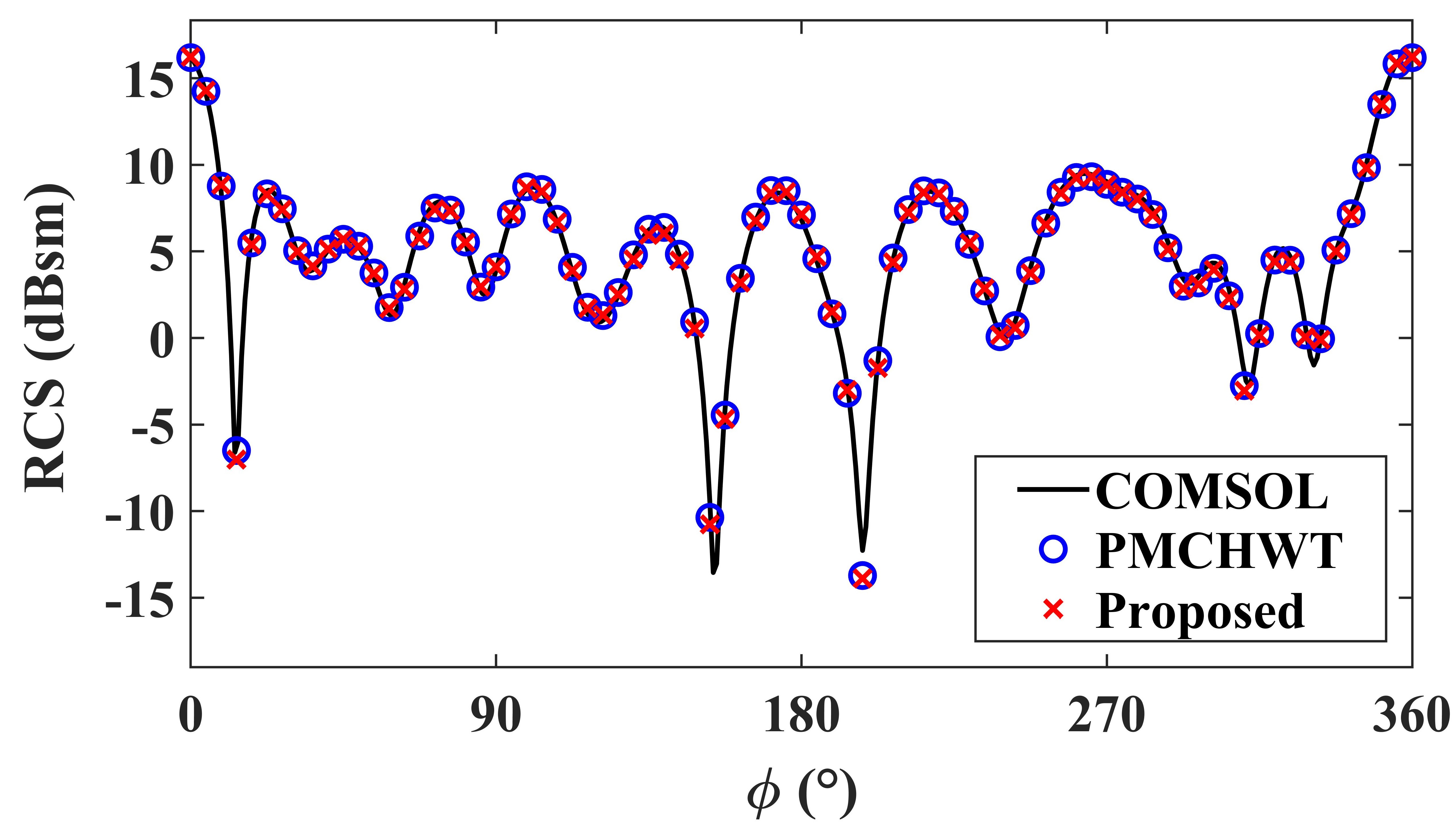}
	\caption{The RCS obtained from the COMSOL, the PMCHWT formulation, and the proposed SS-SIE formulation.}
	\label{R4}
\end{figure}
Fig. \ref{R4} shows the RCS obtained from the COMSOL, the PMCHWT formulation, and the proposed SS-SIE formulation. It is easy to find that the results from the three approaches are again in good agreement. To finish this simulation, the PMCHWT formulation uses 21.0 seconds and 1,406 unknowns, where the averaged length of segments is $\lambda_0/30$ in the conformal meshes. However, the proposed SS-SIE formulation only takes 509 unknowns and { 7.9} seconds, since nonconformal meshes are used and only the electric current density is required.

\begin{figure*}
	\begin{minipage}[h]{0.3\linewidth}
		\centerline{\includegraphics[scale=0.07]{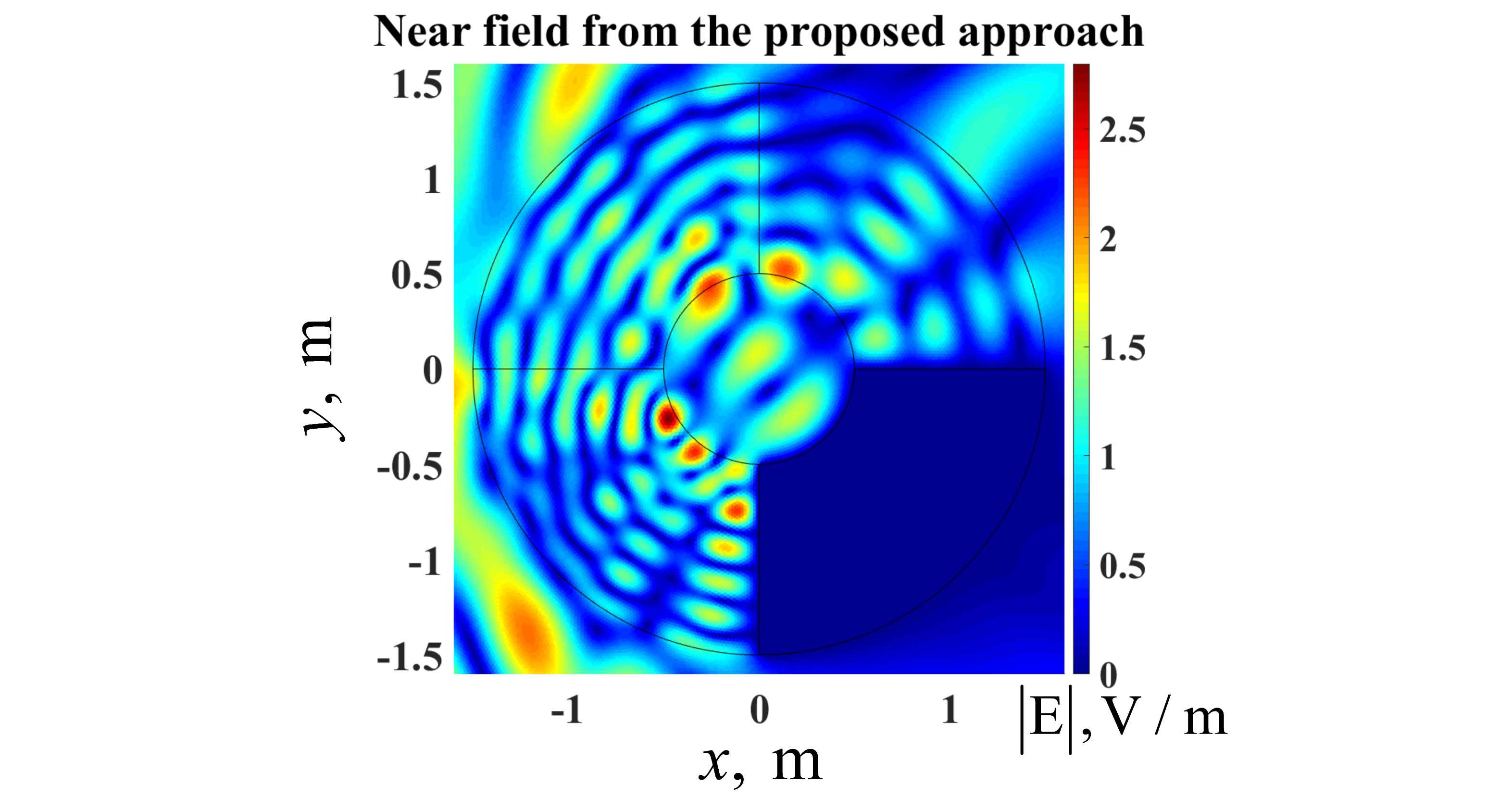}}
		\centerline{(a)}
	\end{minipage}
	\hfill
	\begin{minipage}[h]{0.3\linewidth}
		\centerline{\includegraphics[scale=0.07]{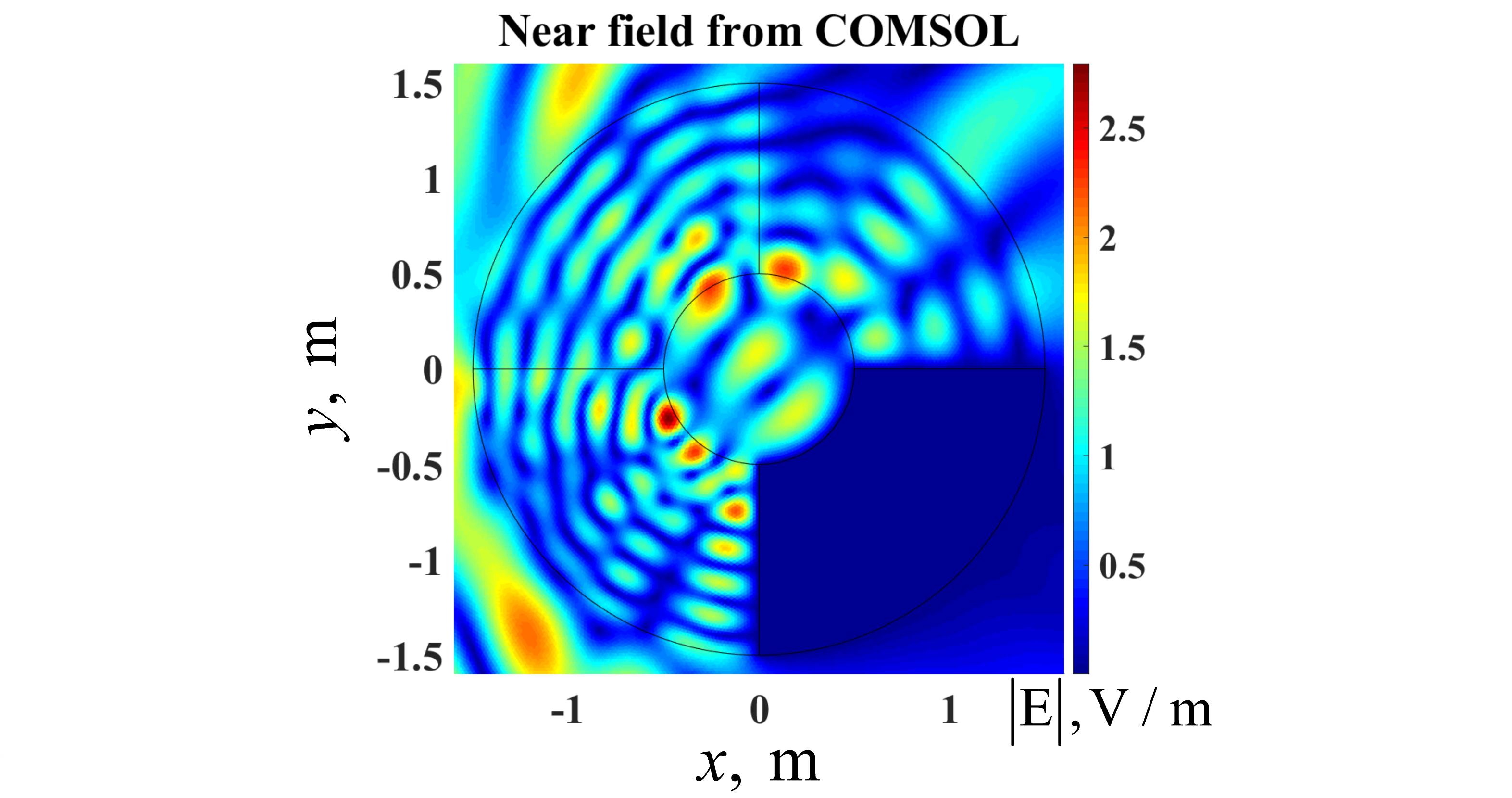}}
		\centerline{(b)}
	\end{minipage}
	\hfill
	\begin{minipage}[h]{0.3\linewidth}
		\centerline{\includegraphics[scale=0.07]{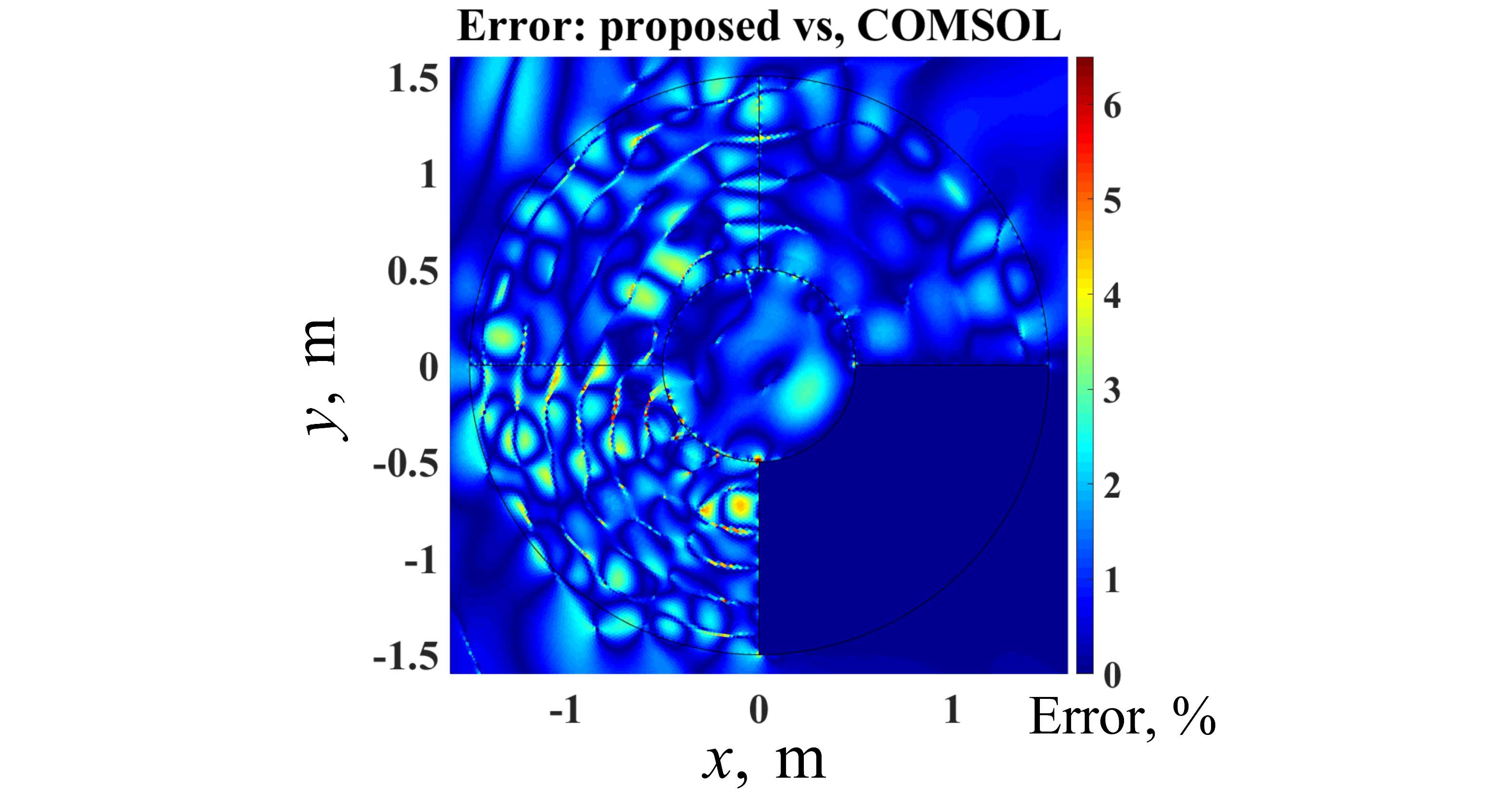}}
		\centerline{(c)}
	\end{minipage} 
	\caption{Near fields obtained from (a) the proposed SS-SIE formulation, (b) the COMSOL, (c) the relative error of near fields obtained from the proposed SS-SIE formulation.}
	\label{N4}
\end{figure*}

The near fields obtained from the proposed SS-SIE formulation and the COMSOL are shown in Fig. \ref{N4}(a) and (b). It can be found that the fields obtained from the two approaches show good agreement. In addition, the relative error of near fields between the two approaches is illustrated in Fig. \ref{N4}(c). As shown in the numerical results, the relative error is less than 4$\%$. Therefore, the proposed SS-SIE formulation is accurate to calculate complex composite structures with partially connected penetrable and PEC objects.

\subsection{Discussion}
As shown in the four numerical examples, the potential of the proposed SS-SIE formulation is demonstrated by modeling simple dielectric objects with non-smoothing boundaries, the electrically large lossy objects, the planar layered objects and the partially connected dielectric and PEC objects. Both near and far fields are accurately and efficiently calculated from the proposed SS-SIE formulation. Although our current implementations are in the TM mode, and our numerical examples are simplified from practical engineering problems, like metallic traces in integrated circuits, the planar multilayer media, the proposed SS-SIE formulation shows great potential in solving practical engineering problems.

{ When electrically large objects are involved in the computational domain, the DSAO-based formulations suffer from efficiency degradation since explicit matrix inversion is unavoidable to construct the single electric current density. The proposed SS-SIE formulation can alleviate this issue. By decomposing the original large structures into small units, the dimensions of matrix requiring to be inverted can be much reduced. However, since additional unknowns will be added on the shared boundaries, the dimension of the final matrix equation becomes larger. This issue can be mitigated if the iterative algorithms along with acceleration methods, such as the MLFMA [\citen{MLFMA}], pFFT [\citen{pFFT}], AIM [\citen{AIM}] are used. However, the proposed SS-SIE formulation may be not suitable to model objects with large circular cross sections since many additional unknowns have to be added with domain decomposition. This issue can be mitigated through the fast direct solver, such as the H-matrix [\citen{JiaoDan-HMatrix}], HSS methods [\citen{S.Ambikasaran-HSS}], and so on.

Another issue we observed in the proposed SS-SIE formulation is resonance since only the EFIE formulation is used to derive the DASO. For lossy media, it is not an issue. However, when lossless dielectric objects are involved and wideband frequency sweep, it has to be handle with the combined field integral equation (CFIE) [\citen{DSAOARBSHAPEDO}] to avoid resonance. In addition, the proposed SS-SIE formulation may also suffer from accuracy issues under conditions of low frequencies, dense materials, and oversampling. It will be a topic in the future.
}

\section{Conclusion}
An efficient and simple SS-SIE formulation for electromagnetic analysis of arbitrarily connected penetrable and PEC objects is developed. Through modularly constructing the equivalent model incorporating with the DSAO for each penetrable objects, and combining the equivalent current densities and physics current densities on the PEC boundaries, an equivalent model for the composite structures with only the electric current densities is derived. The proposed SS-SIE formulation shows many significant advantages over other existing techniques, like implicitly enforced boundary conditions, easy implementation, intrinsically nonconformal mesh support and only the single electric current density. Those merits are quite useful for challenging electromagnetic simulations, like scattering from multiscale and electrically large objects, parameter extraction for high density integrated circuits. As our numerical results shown, significant performance improvement in terms of the CPU time and memory consumption is obtained compared with the traditional PMCHWT formulation. It is much more flexible than the original SS-SIE formulation to solve the challenging electromagnetic problems. 

Extension of current work into three dimensional general scenarios is in progress. We will report more results upon this topic in the future.

\section*{Appendix}
\begin{figure}[H]
	\centering
	\includegraphics[width=0.322\textwidth]{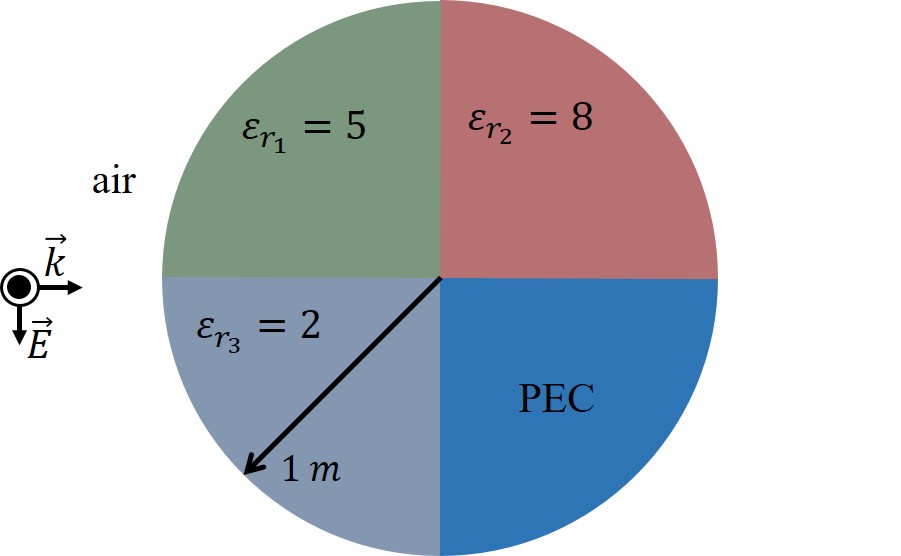}
	\caption{A composite structure consisting of four quarters including three dielectric objects and one PEC object.}
	\label{S5}
\end{figure}

\begin{figure}
	\centering
	\includegraphics[width=0.21\textwidth]{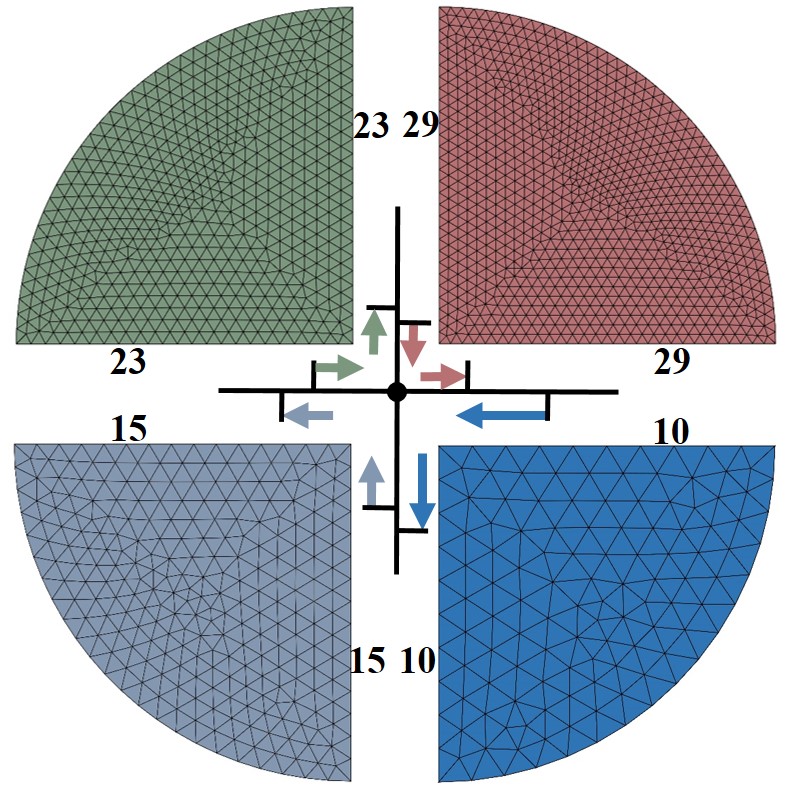}
	\caption{The nonconformal meshes for the composite object, the numbers indicate the count of segments on the shared boundaries used in our simulations. A small gap is added and the triangle edges are presented to better visualize the nonconformal meshes. Only the boundary segments are used in the simulations. }
	\label{S5_m}
\end{figure}

An extensional example is presented to demonstrate the capability of the proposed SS-SIE formulation to solve the vector TE mode in Fig. \ref{S5}. There are three dielectric quarter cylinders with the relative permittivity of $\varepsilon_{r_1} = 5$, $\varepsilon_{r_2} = 8$, $\varepsilon_{r_3} = 2$ and a PEC quarter cylinder. The radii of the concentric quarters is 1 m. The basis function used to discretize the vector electric fields and electric current density is the rooftop basis function, which mimics the RWG basis function in three dimensional space [\citen{MOMBOOK}, Ch. 2]. In Fig. \ref{S5_m}, four basis functions are marked at the junctions of the four quarter cylinder intersection. The averaged mesh size of the four boundaries is $\lambda_0/22$, $\lambda_0/28$, $\lambda_0/14$, $\lambda_0/10$, respectively. It is obvious that the mesh is nonconformal and the count of segments for each region is shown in Fig. \ref{S5_m}. Similar to our previous two numerical examples, a small gap is added and the triangle edges are added to better visualize the nonconformal meshes. In this example, a plane wave with a frequency of 300 MHz incidents from the $x$-axis. 

\begin{figure}
	\centering
	\includegraphics[width=0.45\textwidth]{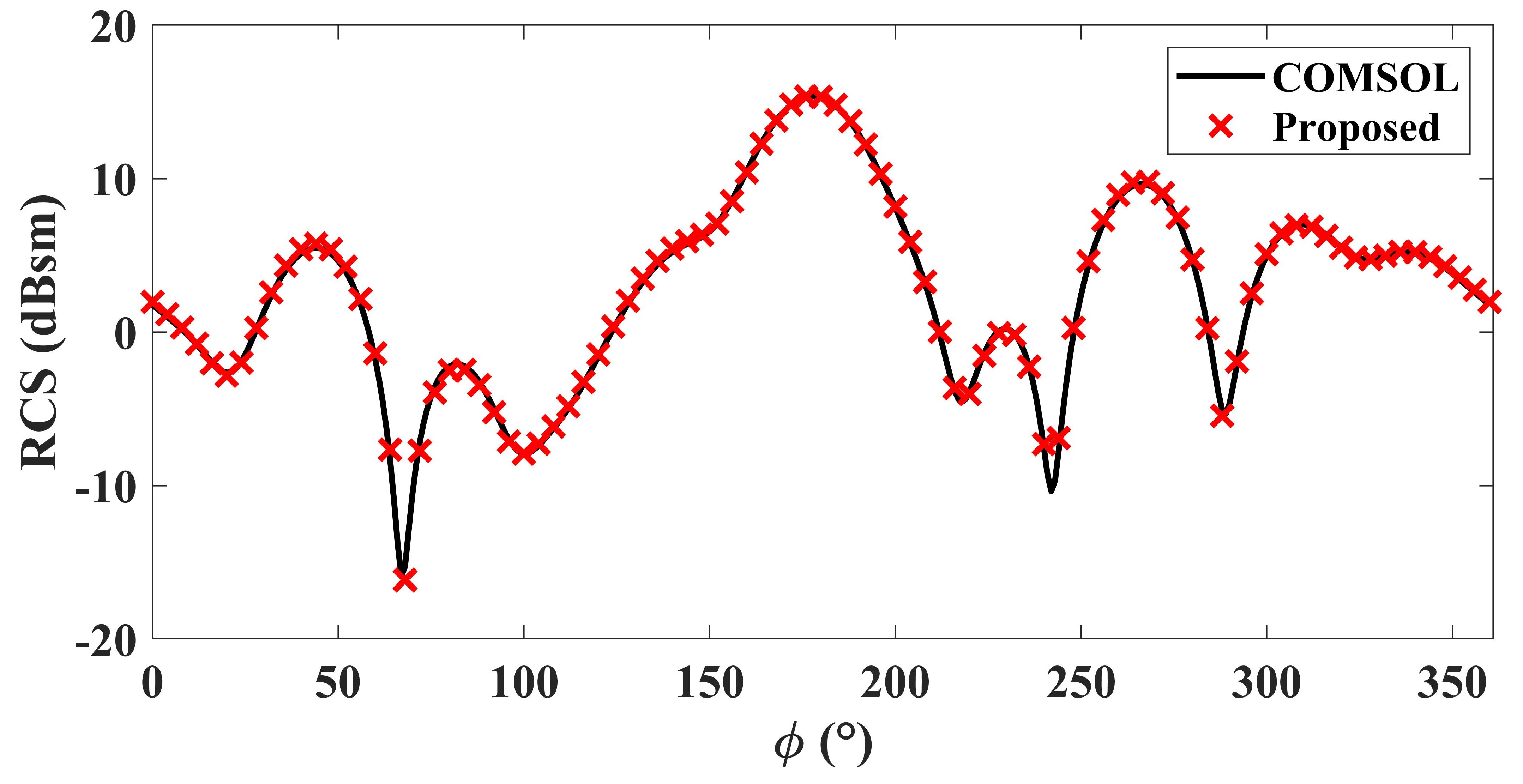}
	\caption{The RCS obtained from the COMSOL and the proposed SS-SIE formulation.}
	\label{R5}
\end{figure}

Unlike the traditional SIE formulations [\citen{putnam1991Combined}]-[\citen{yla2005application}], in which special attention should be paid to the boundary conditions at the junction, the proposed SS-SIE formulation does not need any special treatments at the junction, which is quite easy to handle this structures. Fig. \ref{R5} presents the RCS obtained from the COMSOL and the proposed SS-SIE formulation. Results obtained from the proposed SS-SIE formulation show excellent agreement with those of the COMSOL. Therefore, the proposed SS-SIE formulation is also applicable to solve the general electromagnetic problems in the vector TE mode, which can significantly simplify the implementations on the junctions of multiple media intersection. 

\end{document}